\documentclass[aps,prl,twocolumn,superscriptaddress,floatfix,export]{revtex4-2} 
\usepackage[colorlinks=true,linkcolor=blue,citecolor=blue,urlcolor=blue]{hyperref}
\hypersetup{
    pdftitle={CounterintuitiveSPTMBQC},
}

\usepackage[utf8]{inputenc}
\usepackage[dvipsnames]{xcolor}
\usepackage[titletoc,toc,title]{appendix}
\usepackage{times}
\usepackage{mathtools}
\usepackage{tikz}
\usepackage{amsthm,physics,empheq,tcolorbox,enumerate,comment}
\usepackage{float}
\usetikzlibrary{shapes.geometric, arrows, shapes.multipart}
\usepackage{booktabs, tabularx}

\usepackage[export]{adjustbox}
\usepackage{titlesec,amsfonts,braket,csquotes}
\usepackage{amssymb}

\newtheorem{theorem}{Theorem}

\def\<#1>{\mathinner{\langle#1\rangle}}

\newtheorem{Cor}{Corollary}

\newtheorem{Lemma}{Lemma} 

\newtheorem{Def}{Definition}

\begin{document}

\title{Counter-intuitive yet efficient regimes for measurement based quantum computation on symmetry protected spin chains} 

\date{\today}

\author{Arnab Adhikary}
\affiliation{Department of Physics and Astronomy,
University of British Columbia, Vancouver, Canada}
\affiliation{Stewart Blusson Quantum Matter Institute, University of British Columbia, Vancouver, Canada}

\author{Wang Yang}
\affiliation{School of Physics, Nankai University, Tianjin, China}

\author{Robert Raussendorf}
\affiliation{Department of Physics and Astronomy,
University of British Columbia, Vancouver, Canada}
\affiliation{Stewart Blusson Quantum Matter Institute, University of British Columbia, Vancouver, Canada}

\begin{abstract}
Quantum states picked from non-trivial symmetry protected topological (SPT) phases have computational power in measurement based quantum computation. This power is uniform across SPT phases, and is unlocked by measurements that break the symmetry. Except at special points in the phase, all computational schemes known to date place these symmetry-breaking measurements far apart, to avoid the correlations introduced by spurious, non-universal entanglement. In this work, we investigate the opposite regime of computation where the symmetry-breaking measurements are packed densely. We show that not only does the computation still function, but in fact, under reasonable physical assumptions, this is the most resource efficient mode.

\end{abstract}

\maketitle

\noindent
Computational phases of quantum matter \cite{Bartl1,M1, Bartl}  
are an intriguing notion at the junction of physics and computer science, specifically condensed matter physics and quantum computation. From the perspective of condensed matter physics, they are symmetry protected topologically ordered (SPT) phases of ground states \cite{GW,Wen1,Schuch,Pollm}. From the perspective of quantum computation, they provide computational resources. 

The computational power of SPT ground states is unlocked by  measurement based quantum computation (MBQC) \cite{RB01}, a universal scheme of quantum computation driven by local projective measurements. Those measurements are applied to a suitable entangled initial quantum state, the resource state.

For many SPT phases 
%in spatial dimensions 1 and 2 
it has been established that every ground state in a given phase has the same utility as MBQC resource state \cite{SPTO1,SPTO2,MM2,2Duniv,SOCO, DAM,QCA,DW}---hence the name `computational phase'.

The analysis of MBQC from the perspective of SPT has taught us that entire MBQC schemes---their computational power as well as their operation---are determined by symmetry \cite{Bartl,SPTO1,SOCO}.
In this regard, both halves of MBQC---the resource states and the measurements---warrant comment.

Symmetry protected states possess two types of entanglement. The first type is protected by symmetry. It is constant throughout any given SPT phase, and accounts for the essential way in which all states in the phase are alike. Second, there is residual ``junk'' entanglement \cite{Bartl}. It varies across the phase, accounting for how the individual ground states differ from one another. From the perspective of MBQC, the symmetry-protected entanglement present in SPT ground states is their main asset, and the residual entanglement a formidable obstacle. Indeed, the main difficulty in using SPT ground states as MBQC resources is to guard against the detrimental effects of residual entanglement \cite{Bartl}. Strategies to accomplish this have been found  \cite{MM2, SPTO1}.

Regarding the measurements that drive MBQC, a curious dichotomy arises. While symmetry characterizes and classifies computational power, symmetry breaking is required to achieve it. Namely, the MBQC resource states are invariant under the given symmetry, but measurements can enact non-trivial logical gates only if they are not.\smallskip

\begin{figure}[b]
\includegraphics[width=8cm]{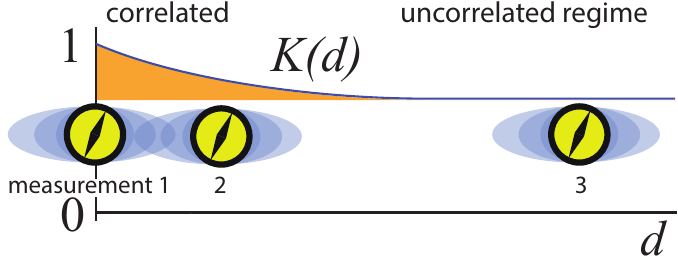}
 \includegraphics[width=7.7cm]{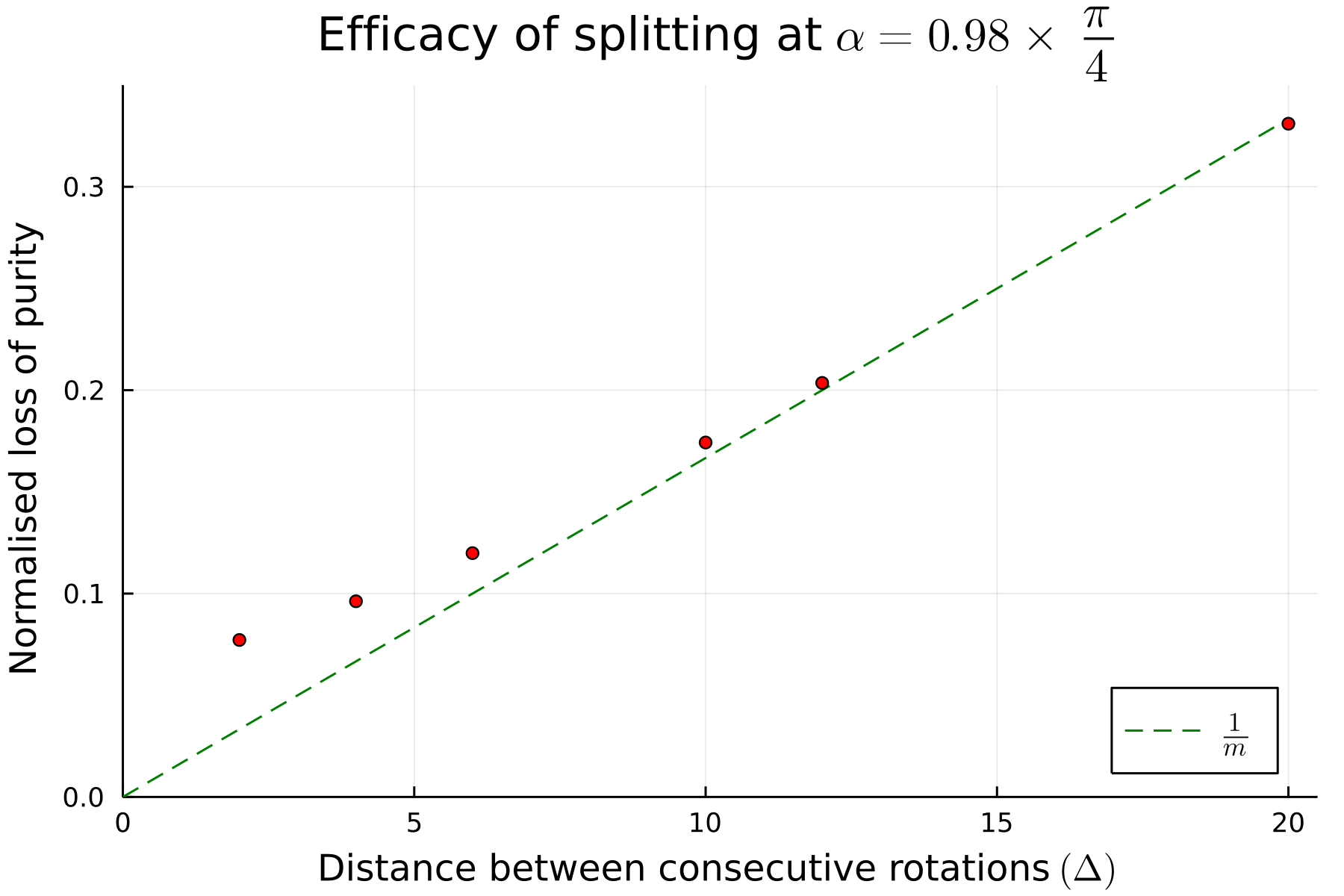}
    \caption{[top] Schematic view of the string order parameter $\mathcal{K}$ vs. distance $d$;  $\mathcal{K}(d)$ is constant for large $d$. If two local measurements are within a distance where $\mathcal{K}$ still changes, they create a logical effect jointly (correlated regime); otherwise they create separate logical effects (uncorrelated regime). [bottom] Numerical simulation of loss of purity vs. density of packing, close to the boundary of the cluster phase. When the packing is already dense, the gains of further increasing the density diminish but remain positive. Densest packing of algorithmically non-trivial measurements is the most efficient.}
    \label{main}
\end{figure}

The present paper is concerned with the question of how densely MBQCs can be packed into a given resource state. A priori, there are two regimes for SPT ground states, the dilute uncorrelated and the dense correlated regime. In the former, the symmetry-breaking local measurements are spaced far apart and in the latter they are close, relative to a characteristic length scale set by the residual entanglement. 

The natural inclination is to stay away from the correlated regime. Indeed, all known ways of performing MBQC on SPT ground states \cite{SPTO1,SPTO2,MM2,2Duniv,SOCO, QCA, DW,DAM} operate in the uncorrelated regime. Therein, the strategy always is to best avoid the effects of  residual entanglement. Up to now, it has not been known whether MBQC is possible in the correlated regime at all. In this paper, we show that it is. Furthermore, under a general condition on the decay of correlations, we establish that the correlated regime is in fact the computationally most efficient.\medskip

\noindent
{\em{The setting: Symmetries and String Order.}} The canonical resource for MBQC is the cluster state. In spatial dimension two and higher it is computationally universal \cite{RB01,RHG}. Here we focus on the simpler one dimensional case, which only encodes a single logical qubit. The 1D cluster state $|\Phi_0\rangle$ on a spin $1/2$ chain of an number $N$  of qubits is a stabilizer state, $K_i|\Phi_0\rangle = |\Phi_0\rangle$, with stabilizer generators $K_i=Z_{i-1}X_iZ_{i+1}$ for $i=2,\ldots,N-1$, and $K_1=X_1Z_2$, $K_N=Z_{N-1}X_N$.
We choose $N$ to be odd.

The 1D cluster state lives inside a symmetry protected topologically (SPT) ordered phase with an abelian symmetry group $G=\mathbb{Z}_2\times \mathbb{Z}_2 $, acting on the spin chain via
\begin{equation}
G \cong \langle g_0, g_1 \rangle =\langle Z_1 X_2 X_4 \ldots  X_{N-1} Z_N, X_1 X_3\ldots X_N\rangle.    
\end{equation}
The ability of the cluster state to transmit quantum information \cite{Bartl} and compute \cite{SPTO1,SPTO2} by local measurement has been extended to the entirety of this phase. 

A one-parameter family of states in the cluster phase is provided by the ground states of the Hamiltonian
\begin{equation}\label{Hamil}
H(\alpha) = -\cos\alpha \sum_{i=1}^N K_i - \sin\alpha \sum_{i=2}^{N-1} X_i, 
\end{equation}
for $|\alpha| < \pi/4$. The cluster state is found at $\alpha=0$. All numerics in this paper are for ground states of $H(\alpha)$.

From a condensed matter viewpoint, the signature of non-trivial SPT order is a non-zero string order parameter \cite{NR}. Here we define such order parameters for all sites separately, not assuming translation invariance, as the expectation values of
\begin{equation}\label{Kdef}
\begin{array}{rcll}
%\mathcal{{K}}_{ \geq k } \\
\mathcal{{K}}_{ \geq k } &:=& Z_kX_{k+1} X_{k+3} .. X_{N-1}Z_N,& \text{for}\; k\; \text{odd},\vspace{1mm}\\
&:=& Z_kX_{k+1} X_{k+3} .. X_N,& \text{for}\; k\; \text{even},\vspace{1mm}\\
%\mathcal{{K}}_{ \leq l } \\
\mathcal{{K}}_{k,l }&:=&Z_kX_{k+1}X_{k+3}.. X_{l-3}X_{l-1}Z_l,&  l-k\; \text{even}.
\end{array}
\end{equation}
 First, we note that if translation  invariance is enforced (because of the exponential decay of correlations, this holds when the couplings in the Hamiltonian are spatially uniform), $\expval*{\mathcal{K}_{\geq }}:= \expval*{\mathcal{K}_{\geq k}}$ for all bulk sites $k$ and $\expval*{\mathcal{{K}}_{k,l }}= \expval*{\mathcal{{K}}{(l-k)}}=:\expval*{{\mathcal{{K}}{(\Delta)}}} $.
We also observe that, as a consequence of the definitions of Eq.~(\ref{Kdef}), the product $\mathcal{{K}}_{ \geq k} \mathcal{K}_{k,l} \mathcal{{K}}_{ \geq l }$, for all choices of $k,l$ both odd or even, is the identity. Therefore, it holds that
$$\langle \mathcal{K}_{\geq k}  \mathcal{K}_{\geq l} \rangle _{\Phi} = \langle \mathcal{K}_{k,l} \rangle_{\Phi}.$$
{ If $\Phi$ is short-range entangled, for large enough separation between the sites $k$ and $l$, the l.h.s. can be further factorized into the product of two expectation values. For a proof of this decoupling see the Appendix Sec I. Finally, assuming translation invariance, we have for large values of $\Delta$ (i.e. in the uncorrelated regime),
$$\ \langle \mathcal{K}({\Delta}) \rangle_{\Phi}=\langle\mathcal{K}_{\geq }\rangle _{\Phi} ^2.$$
}
\noindent
{\em{Results.}}
The central ingredient for MBQC in SPT phases is the splitting of operations into $m$ smaller parts. In the uncorrelated regime, for the logical operation $T(\beta)$ evoked by the measurement an angle $\beta$ away from the symmetry-respecting local basis (the $X$-basis in our case) the following has been shown \cite{SPTO1,SPTO2}: (i) the deviation of $T(\beta)$ from the identity is linear in $\beta$, and (ii) the deviation of $T(\beta)$ from unitarity is quadratic in $\beta$. Therefore, it is of advantage to split $T(\beta)$ into $m$ operations $T(\beta/m)$. In this way, the overall error of the operation scales as $1/m$, and can thus be made arbitrarily small. The splitting of operations ensures that the same computational power is maintained all across a given SPT phase.

To quantify the logical error $\mathcal{D}$ incurred in the process of the computation, we define it as the trace distance between the intended and implemented logical operation (for details see the Appendix Sec. III).

\begin{theorem}\label{T1}
Throughout the 1D cluster phase with $\mathbb{Z}_2 \times \mathbb{Z}_2$ symmetry, for all separations $\Delta$ between successive symmetry-breaking measurements, the logical error $\epsilon$ for a $x$- or $z$-rotation about an angle $\beta_{\text{log}}$ can be made arbitrarily small, by splitting the rotation into $m$ parts. The logical error is $\mathcal{D}_m = \beta_{\text{log}}^2 \,\kappa/m+ O(1/m^2) $, 
with
\begin{equation}\label{kappa}
\kappa = \left(1+2\sum_{j=1}^{m-1}f(j\Delta)\right)\left(\frac{1}{\expval*{\mathcal{K}_{\geq}}^2}-1\right)\;
\end{equation}
a constant depending on the string order parameter. Therein, $f(\Delta):= (\expval*{\mathcal{K}{(\Delta)}}-{\expval*{\mathcal{K}_{\geq}}}^2 )/(1-{\expval*{\mathcal{K}_{\geq}}}^2)$.
\end{theorem}

Herein, the function $f$ characterises the range of correlations. In all cases, $f(0)=1$ and $f \longrightarrow 0$ at large distances. The correlated regime applies if $f(\Delta)$ appreciably deviates from zero. In the uncorrelated regime it holds that $f(j\Delta)=0$ for all $j\in \mathbb{N}$, and $\kappa$ simplifies to $(1-\expval*{\mathcal{K}_{\geq}}^2 )/\expval*{\mathcal{K}_{\geq}}^2 $, reproducing Corollary 1 in \cite{SOCO}.

\begin{figure}
\begin{center}
\includegraphics[width=8cm]{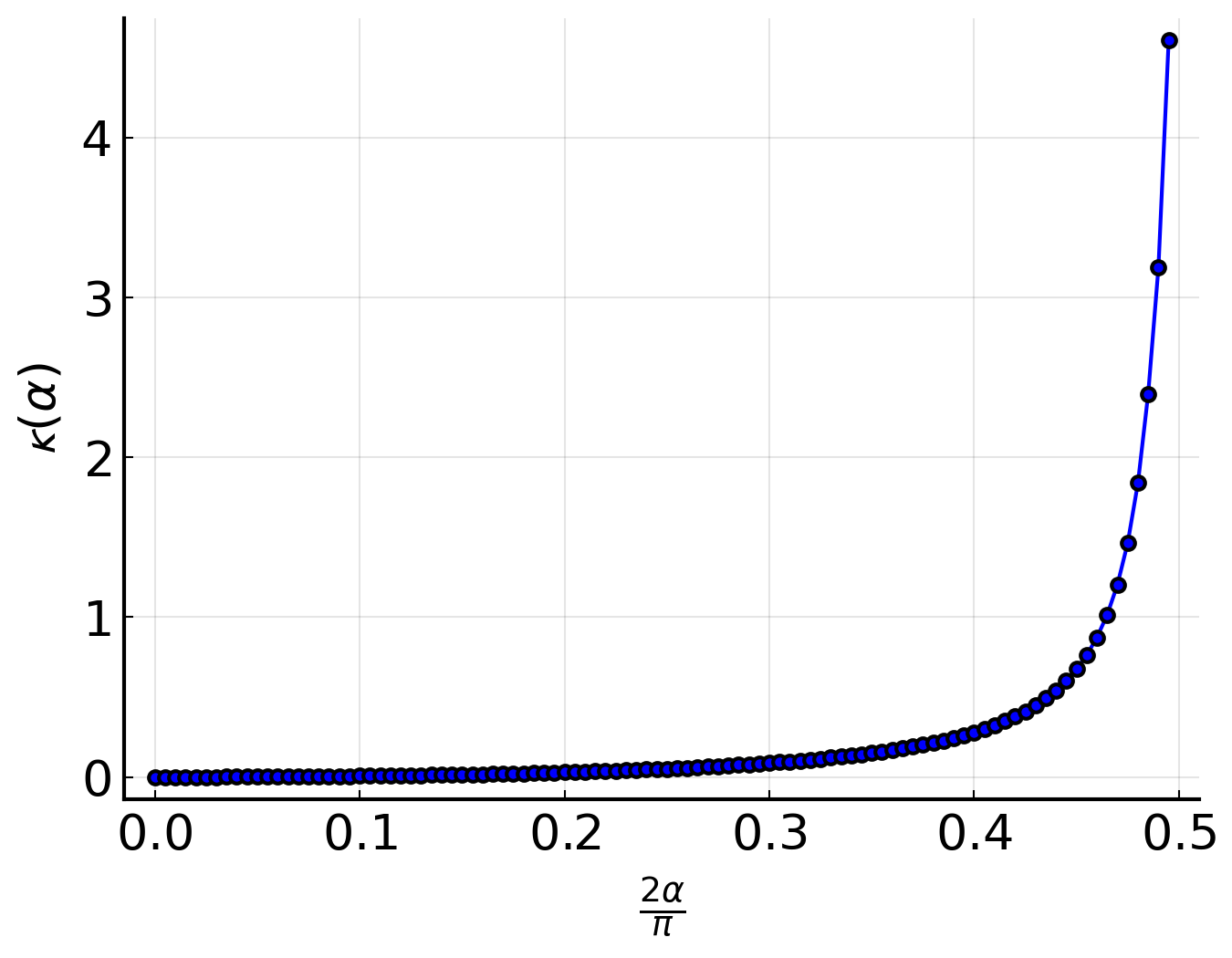}
    \caption{\label{kappaFig}
     The efficiency characteristic $\kappa$ of resource states in the cluster phase, vs. the location $\alpha$ within the phase. The operational cost to implement a given unitary logical gate with target error $\epsilon$ is proportional to $\kappa/\epsilon$; cf. Theorem~\ref{T1}. The value of $\kappa$ is small throughout most of the cluster phase, raising only near the phase boundary. For the cluster state itself ($\alpha=0$), it holds that $\kappa=0$. This means that at the cluster point, unitary gates can be realized perfectly. For this plot, we chose $\Delta=2$, in anticipation of Theorem~\ref{T2}. (Numerics are based on a chain of length $N=501$.)}
\end{center}
\end{figure}

Regardless of the strength and range of correlations, the same scaling ${\cal{D}}_m \propto  \kappa/m$ of the logical error vs. the invested computational resources applies. Herein, the parameter $\kappa$ varies across the phase. Thus, while all states in the cluster SPT phase can be used for the same quantum computations, some are more efficient than others. For the perfect resource state, the cluster state, it holds that $\kappa=0$, and for a poor resource state near the phase boundary, $\kappa\gg 1$; see Fig.~\ref{kappaFig}. \smallskip

If the same scaling law holds for both the uncorrelated and the correlated regime, then which one is better? In this regard, we provide the following result.

\begin{theorem}\label{T2}
In MBQC on any state in the $\mathbb{Z}_2 \times \mathbb{Z}_2$ 1D cluster phase for which the string order parameter $\expval*{{\cal{K}}(d)}$ is a convex function of the distance $d$, the densest packing of symmetry-breaking measurements, $\Delta = 2$, is the most computationally efficient.
\end{theorem}

Thus, assuming the condition holds, it is not the optimal strategy to avoid correlations stemming from the residual entanglement, but rather to adjust to them.

We observe that the precondition of convex decay of $\expval*{{\cal{K}}(d)}$ appears to be broadly satisfied. Indeed, in all our numerical simulations, no deviation from convexity is found. It also seems to be ``natural'' in the following sense: Irrespective of the model under consideration, the string order parameter (i) starts out at its maximum possible value of $1$ and (ii) settles to a constant for large values of the string length. The simplest way to satisfy both constraints is by a convex function.\smallskip  

The above Theorems are proved in the Appendix Sec III \& IV. Below we analyze the simplest instance to which they apply, MBQCs with two symmetry breaking measurements.\smallskip

Before discussing the example, we review required background on MBQC. The need to account for the randomness in the outcomes of the measurements driving the computation, gives rise to the byproduct operators \cite{RB01}. The following facts are known: (i) The byproduct operators are constant throughout the phase \cite{Bartl}. (ii) The byproduct operators are $X^s$ on even sites and $Z^s$ on odd sites, with $s\in \mathbb{Z}_2$ the measurement outcome for the respective sites \cite{RB01}. (iii) The local measurement on spin $k$ is in the eigenbasis of $\cos \beta X_k -(-1)^{q_k} \sin\beta Y_k$, where $q_k\in \mathbb{Z}_2$ depends on measurement outcomes to the left, $q_k=s_{k-1}+s_{k-3}+..+s_{2/1}$ \cite{RB01}.

Denote the $\mathbb{Z}_2\times \mathbb{Z}_2$-symmetric resource state by $|\Phi\rangle$. Then, with facts (i) and (ii), the logical expectation values of the single logical qubit processed are
\begin{equation}\label{BasExp}
\begin{array}{rcl}
\langle {\cal{X}} \rangle &=& \langle \Phi| O_1 O_3 O_5.. O_{N-2} X_N |\Phi\rangle,\\
\langle {\cal{Z}} \rangle &=& \langle \Phi| O_2 O_4 O_6 .. O_{N-1} Z_N |\Phi\rangle,\\
\langle {\cal{Y}} \rangle &=& \langle \Phi| O_1 O_2 O_3 .. O_{N-1} Y_N |\Phi\rangle,
\end{array}
\end{equation}
Herein, the observables $O_k$ represent the above $\cos \beta X_k -(-1)^{q_k} \sin\beta Y_k$, but take all adaptation of measurement basis into account at the quantum mechanical level. That is,
$O_k = \sum_{\textbf{s}}(-1)^{s_k}|\textbf{s}\rangle \langle \textbf{s}|$,
where the states $|\textbf{s}\rangle$ form the local MBQC measurement basis,
$|\textbf{s}\rangle = \bigotimes_{i=1}^N |s_i, q_i(\textbf{s}_{\prec i})\rangle$,  $\textbf{s}_{\prec i}$ is the measurement record acquired prior to the measurement at $i$, and $q_i \in \mathbb{Z}_2$ decides the local measurement basis at site $i$.

To form an intuition, one may consider MBQC as a two-step process in which, first, a one-qubit state is prepared on the rightmost spin of the chain, and, second, this last spin is subjected to state tomography, requiring the measurement of $X$, $Y$ and $Z$. Wlog, we assume the chain to be of odd length.
From this perspective, the reason why $\langle {\cal{X}} \rangle$ /  $\langle {\cal{Z}} \rangle$ /  $\langle {\cal{Y}} \rangle$ in Eq.~(\ref{BasExp}) include the observables on odd / even / all sites $<N$ is that the final (tomographic) measurement of $X$ / $Z$ / $Y$ is affected by byproduct operators $Z$ (generated on odd sites) / byproduct operators $X$ (generated on even sites) /  both types of byproduct operators.\medskip

\noindent
{\em{Example.}} To illustrate Theorem~\ref{T1}, here we analyze the working of MBQC in the correlated regime for the simplest instance that exhibits correlations. This occurs with two symmetry-breaking measurements, $m=2$. To further simplify, we fix the input state of the logical qubit to $|+\rangle$, which happens naturally in the cluster chain. In this setting, we can do the required derivations from scratch in a compact fashion, rather than appealing to the theorems. The basic insights can be verified in this simple scenario.\smallskip

The example MBQC simulates the following circuit: (I) Preparation of $|+\rangle$, (II) logical operations, (III) State tomography on the output qubit. The logical operations in step II are implemented  on sites $k$ and $l$ (both odd), and invoke measurements at an angle $\beta_k$ and $\beta_l$ away from the symmetry-respecting $X$-basis. On the cluster state as computational resource, these measurements implement logical rotations about the $z$-axis, with rotation angles $\beta_k$ and $\beta_l$, respectively. The question is: Throughout the cluster phase, which logical operations are generated, and how much purity is lost?\smallskip

The observables in Eq.~(\ref{BasExp}) simplify in this example; e.g. $\langle {\cal{X}} \rangle = \langle X_1 X_3 ..O_k X_{k+2} .. O_l.. X_N\rangle$. Further, by fact (iii) above, the observables $O_k$ and $O_l$, with all adaptation included, can be written as
\begin{equation}\label{Ok}
\begin{array}{rcl}
O_k &=& \cos \beta_k X_k - \sin \beta_k X_2 X_4..X_{k-1} Y_k,\\
O_l &=& \cos \beta_l X_l - \sin \beta_l X_2 X_4..X_{l-1} Y_l.
\end{array}
\end{equation}
Inserting this into the above expressions for the three expectation values $\langle {\cal{X}} \rangle$,  $\langle {\cal{Y}} \rangle$,  $\langle {\cal{Z}} \rangle$, we get four terms in each case. They are either a symmetry, a string order parameter, or anti-commute with a symmetry.  We obtain
\begin{equation}\label{effects}
\begin{array}{rcl}
\langle {\cal{X}} \rangle &=& \cos \beta_k \cos\beta_l-\sin\beta_k \sin\beta_l \expval*{\mathcal{K}_{k,l}},\\
\langle {\cal{Y}} \rangle &=& \sin \beta_k \cos \beta_l \expval*{\mathcal{K}_{\geq{k}}}+ \sin \beta_l \cos \beta_k \expval*{\mathcal{K}_{\geq l}} \\
\langle {\cal{Z}} \rangle &=&  0.
\end{array}
\end{equation}
As a consistency test, for $\beta_k=\beta_l=0$, this amounts to the transmission of a logical qubit in state $|+\rangle$ and its subsequent measurement. Eq.~(\ref{effects}) provides $\langle {\cal{X}}\rangle=1$ throughout the cluster phase, in agreement with \cite{Bartl}.\smallskip

It turns out that as long as the  initial logical state is fixed to $|+ \rangle$, the decoherence measure ${\cal{D}}$ coincides with $\mathcal{D}= 1 -{\expval*{\mathcal{X}}}^2-\expval*{\mathcal{Y}}^2-\expval*{\mathcal{Z}}^2$ up to leading order in the local measurement angles.
We now specialize to two cases, the non-split and the split rotation.\smallskip 

{\em{Case 1,}} $\beta_k=\beta, \beta_l=0$. Eq.~(\ref{effects}) gives $\langle {\cal{X}} \rangle = \cos \beta$, $\langle {\cal{Z}} \rangle =0$, and  $\langle {\cal{Y}} \rangle =  \sin \beta \langle {\cal{K}}_{\geq k} \rangle_\Phi$. To leading order in the measurement angle, such an operation corresponds to a rotation of the logical qubit about the $z$-axis by a reduced angle 
${\beta}_{\text{log}} \approx \beta \langle \mathcal{K}_{\geq k}\rangle$,
with a loss of purity
\begin{equation}\label{Case2}
\mathcal{D}_1= \beta^2 \left(1-\langle \mathcal{K}_{\geq k } \rangle^2\right)+O(\beta^3). 
\end{equation}

{\em{Case 2,}} $\beta_k=\beta_l=\beta/2$.  Since both measurement angles are non-zero, two  non-trivial logical operations are applied. The factor of 1/2 in the measurement angles represents the splitting. Eq.~(\ref{effects}) yields $\langle {\cal{Z}} \rangle=0$ and
$$
\begin{array}{rcl}
\langle {\cal{X}} \rangle &=& \cos^2(\beta/2)-\sin^2(\beta/2) \langle {\cal{K}}_{ k,l}\rangle \\
\langle {\cal{Y}} \rangle &=& \sin (\beta/2) \cos (\beta/2) \left(\langle{\cal{K}}_{\geq k} \rangle + \langle {\cal{K}}_{\geq l} \rangle \right),
\end{array}
$$
To leading order in $\beta$, the logical qubit is rotated about the $z$-axis by an angle
$\beta_{\text{log}}= \beta\frac{\langle {\mathcal{K}_{\geq k}} \rangle +\langle {\mathcal{K}_{\geq l}}\rangle}{2} + O(\beta^3)$. Assuming translation invariance, $\langle {\mathcal{K}_{\geq k}} \rangle = \langle {\mathcal{K}_{\geq l}}\rangle$, this is the same operation as in Case~1. We note in particular that the logical rotation performed is, to leading order,  independent of the distance between two symmetry breaking measurement sites. This is a first indication that the ability to perform MBQC extends to the correlated regime.

The loss in purity is 
 \begin{equation}\label{D2}
\mathcal{D}_2 \;
   = \beta^2\;\left(1-\expval*{\mathcal{K}_{\geq}}^2\right) \;\frac{1+f(\Delta)}{2}+ O(\beta^3).
\end{equation}
The scaling of the loss in purity remains quadratic in the measurement angle $\beta$, which suggests that splitting of operations continues to work in the correlated regime. 

Specifically, in the uncorrelated regime, $f(\Delta)=0$, comparing with Eq.~(\ref{Case2}) we observe that the amount of decoherence is cut in half, exactly as expected for splitting into two operations.
Further, except for the extreme long-range correlated case, $f(\Delta)=1$, there is  reduction in logical decoherence. This confirms Theorem~\ref{T1} in the simplest instance.

 \begin{figure}
    \centering
    \includegraphics[width=0.45\textwidth]{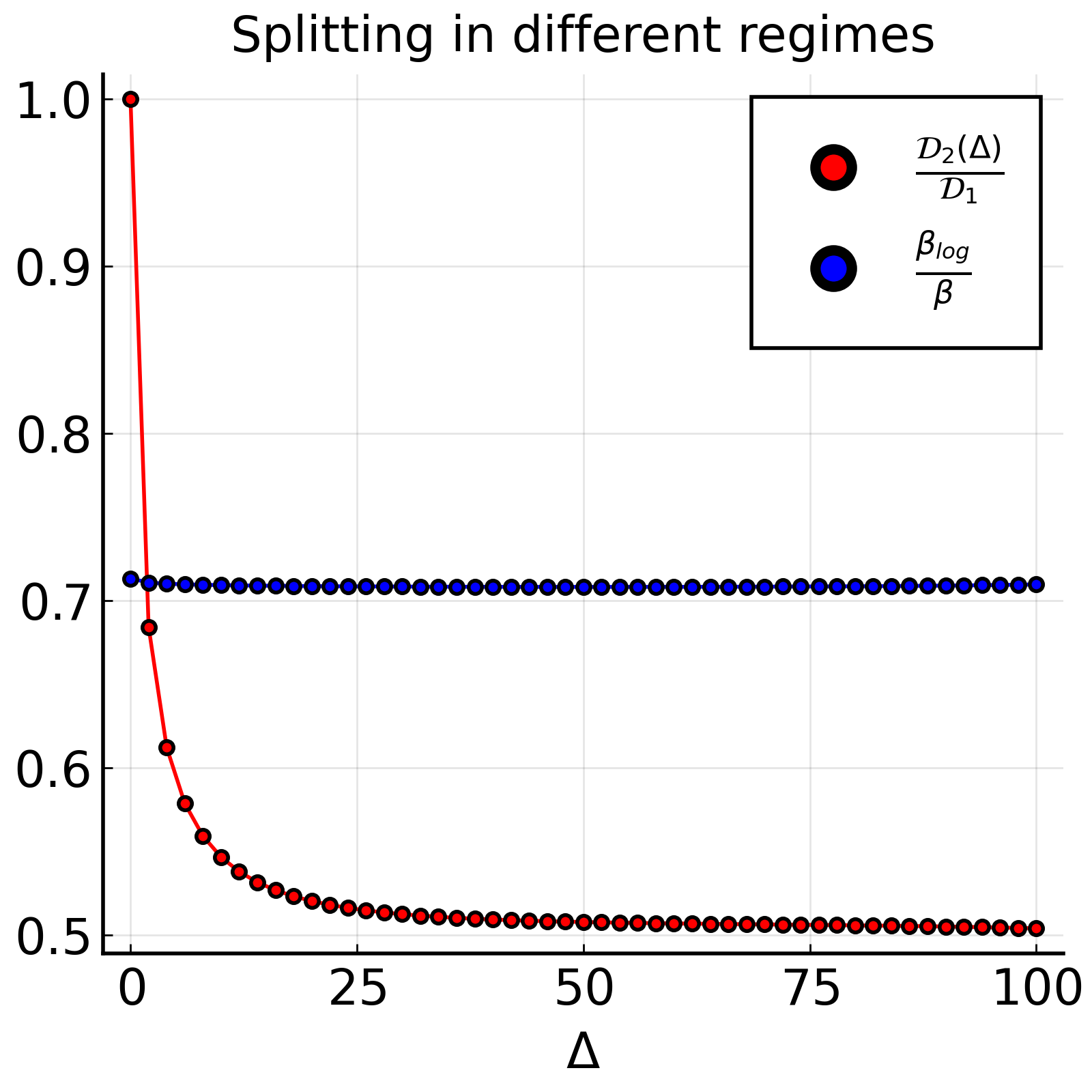}
    \caption{Splitting at work: Reduction in the normalised decoherence metric (see inset) as a function of the distance $\Delta$ between two rotation sites for the ground state of $H(\alpha=0.98 \times \frac{\pi}{4} )$ supported on a chain length of $N=201$. Since the ground state is chosen very close to the phase transition, the length scale of decay is significant. Finally, from the figure we see that the string order decays a convex function of its length which is a prerequisite for Theorem~\ref{T2} to hold. }
    \label{Fig:2site divide}
\end{figure}

 A numerical simulation of the effect of splitting is shown in Fig.~\ref{Fig:2site divide}. Therein, the normalized loss in purity ($=1$ for not splitting) is plotted against the separation $\Delta=l-k$. The simulation is for the ground state of $H(\alpha)$ of Eq.~(\ref{Hamil}) near the phase transition where the correlation length is large, $\alpha= 0.98\times \pi/4$. We find that for large separations $\Delta$, the relative loss of purity drops to 1/2, in agreement with the prediction for splitting. But furthermore, for all values $\Delta>0$ splitting increases purity. 

Yet the figure also shows that the deeper into the correlated regime, the smaller the gain in purity. Thus, when reducing the separation $\Delta$ between two successive symmetry breaking measurements, there are two competing effects, namely (a) splitting becomes less effective, and (b) more splitting can be fit into the same space. Which effect dominates?--This is clarified by Theorem~\ref{T2} above. 

A numerical confirmation is obtained for the ground state of $H(\alpha)$ of Eq.~(\ref{Hamil}) at $\alpha = 0.98\times \pi/4$, see the lower panel of Fig.~\ref{main}. We find that the loss in purity is the lowest for the smallest possible value of $\Delta$, i.e., for $\Delta =2$.\smallskip

Before concluding, we give a glimpse into the proof of Theorem~\ref{T1}. The central tool is the product formula Eq.~(26) in Appendix II. For the present case of fixed logical input $|+\rangle$, and splitting the $z$-rotation into $m$ equal parts, the expectation values of the logical Pauli observables at readout are 
\begin{equation}\label{PF}
%\begin{align*}
\begin{array}{rcl}
  \expval*{\mathcal{X/Y}} &=& \displaystyle{ \Re / \Im \biggl<{\prod_{k=0}^{m-1} \exp
  \left(i\gamma \mathcal{K}_{\geq k\Delta}\right)}\biggr>  ,} 
%\end{align*}
\end{array}
\end{equation}
with $\gamma=\beta_{\text{log}}/(m\expval*{\mathcal{K}_{\geq}})$ the measurement angle. The reader is invited to reproduce the expectation values $\langle{\cal{X}}\rangle$, $\langle{\cal{Y}}\rangle$ for Cases 1, 2 of the Example from the expressions in Eq.~(\ref{PF}).

In the uncorrelated regime, the expectation value in Eq.~(\ref{PF}) can be taken inside the product. The upshot of Theorem \ref{T1} is that despite this crucial difference, efficiency of the splitting strategy persists in the  correlated regime.
\medskip

{\em{Conclusion and Discussion.}} 
In this paper we have demonstrated that measurement-based quantum computation on resources states from a symmetry protected topologically ordered phase can access the correlated regime, where the symmetry-breaking measurements that drive the computation are densely packed. In this correlated regime, the individual local measurements do not individually generate logical operations. Rather, multiple consecutive measurements generate a logical operation jointly, and this operation cannot be decomposed into parts attributed to individual measurements. Surprisingly, this more complicated correlated regime  is generally the most computationally efficient.

\smallskip

The present analysis only covers the $\mathbb{Z}_2 \times \mathbb{Z}_2$ cluster phase in 1D, but we expect our results to broadly generalize. We conclude with two conjectures.

{\em{Conjecture 1.}} Theorems~\ref{T1} and \ref{T2} apply to SPT phases with larger symmetry groups of type $(\mathbb{Z}_2)^k$ in spatial dimensions one and higher, including the computationally universal scenarios \cite{DTS} in 1D and \cite{2Duniv,DW,QCA,DAM} in 2D.

In support of this conjecture we observe that our above results only depend on the string order parameter $\langle {\cal{K}}(d)\rangle$ as a function of distance $d$. Computation is efficient if the order parameter differs from zero substantially, and settles into its long-distance limit quickly. Whether the order parameter is string-like, as in 1D SPT phases, or cone-like with an internal checkerboard or fractal pattern, as in 2D SPT phases with large symmetry groups \cite{2Duniv,DW,QCA,DAM}, seems  immaterial.\smallskip

Symmetry protected topologically ordered states are short-range entangled \cite{Wen1}. This property is not necessary for MBQC; as the toric code example demonstrates \cite{MBQCtc}.

{\em{Conjecture 2.}} The characterization of the efficiency of an MBQC in terms of the single parameter $\kappa$, as in Theorem~\ref{T1}, extends beyond SPT to topological order and symmetry enriched topological order.\footnote{We remark that the present proof of Theorem~\ref{T1} uses short-range entangledness as a background assumption, but this is only to simplify the theorem statement. Namely, it speaks about simulating finite $x$- and $z$-rotations, rather than executing one computation as a whole.}.

In support of this conjecture, we turn to the topologically ordered Kitaev surface code state \cite{Kita,BK} as an MBQC resource. Its computational power is limited by the mapping to non-interacting Majorana fermions  or matchgates \cite{Barahona, Val,MatchJo}. Nonetheless, the surface code state gives rise to a perfectly executable MBQC scheme \cite{MBQCtc}. The condition in Theorem~\ref{T1} recognizes this, $0$ for the toric code state, w.r.t. the order parameters that are computationally relevant (See Fig.~4 in Appendix V). We anticipate that in the general case the expression for $\kappa$ extracts the long-range effects that are harmful to MBQC, separating out those that are not.\smallskip

To summarize, the discussion of computational phases of quantum matter has an algebraic and a non-algebraic part. The former clarifies what is MBQC-computable in a given physical phase with symmetry, and the latter describes how efficient that computation is, for any fixed resource state in the phase. 

The two theorems of this paper settle efficiency for the 1D SPT cluster phase, and Conjectures 1 \& 2 provide avenues for generalization. The algebraic classification of computational phases of quantum matter, on the other hand, is wide open.
\smallskip

{\em{Acknowledgments.}} This work is funded by NSERC, in part through the Canada First Research Excellence Fund, Quantum Materials and Future Technologies Program, and by USARO (W911NF2010013). The numerical simulations in this work were performed using the software package ITensor \cite{itensor}. AA thanks E.M. Stoudenmire for software support.\smallskip

W.Y. and A.A. contributed equally to this work.
\appendix

\bibliography{main.bib}

\appendix

\onecolumngrid

\begin{appendices}

\section{Appendix: Counter-intuitive yet efficient regimes for measurement based quantum computation on symmetry protected spin chains}

The purpose of this appendix is to prove Theorems~1 and 2 of the main text. It is organized as follows. In Section~\ref{SRE} we review the applicable notion of `short-range entanglement'. In Section~\ref{sec:product_formula} we derive a product formula representing the sequence of logical operations, cf. Eq.~(\ref{FactoredPaulis}). It is central for the subsequent proofs. In Section~\ref{PT1} we prove Theorem~1, and in Section~\ref{PT2} Theorem 2. Finally, Section~\ref{SC-MBQC} displays the surface code state as computational resource, as illustration for Conjecture 2.

\setcounter{secnumdepth}{1}

\section{:  Short-range entangled states}\label{SRE}

We first briefly review the definition of short-range entangled states in Ref. \cite{SOCO} which is relevant to the resource states that we consider in this paper.
The resource states for measurement based quantum computation (MBQC) are all of the form 
\begin{eqnarray}
|\Phi\rangle = U_\Phi (|+\rangle |+\rangle..|+\rangle),
\end{eqnarray}
in which $U_\Phi$ is a  bounded-depth circuit composed of bounded-range gates.
Denote $\{\leq k\}$ and $\{>k\}$ to be 
\begin{eqnarray}
\{\leq k\}:= \{0,1,2,..,k\},\; \{>k\} :=\{k+1,k+2,..,N\},
\end{eqnarray}
and  $\text{supp}(A)$ to be the support of a linear operator $A$.
We have the following definition.
\begin{Def}\label{def:SRE_states}
$U_\Phi$ is a short-range entangled quantum circuit with entanglement range $\xi$ if 
\begin{equation}\label{DefRangeEq}
\begin{array}{rl}
\text{supp}(U^\dagger_\Phi AU_\Phi) \subset \{\leq (k+\xi)\},& \forall \,A|\, \text{supp}(A) \subset \{\leq k\},\\
\text{supp}(U^\dagger_\Phi AU_\Phi) \subset \{>(k-\xi)\},& \forall \,A|\, \text{supp}(A) \subset \{> k\}.
\end{array}
\end{equation}
\end{Def}

Here we note that the symmetry protected topological (SPT) states in the cluster phase considered in this work do not exactly satisfy Eq. (\ref{DefRangeEq}).
However, the difference from Eq. (\ref{DefRangeEq}) is only some ``exponentially small tail", i.e., $U^\dagger_\Phi AU_\Phi$ is exponentially small outside of the region $[k-\xi,k+\xi]$. 
Therefore, using the definition in Def. \ref{def:SRE_states} does not do any essential harm to our subsequent analysis, but has the advantage that the discussions can be simplified a lot. 

\begin{Lemma}\label{Prod}
Consider a short-range entangled state $|\Phi\rangle=U_\Phi|+\rangle|+\rangle..|+\rangle$, where the circuit $U_\Phi$ has an entanglement range $\Delta$. Be $A$ and $B$ two linear operators, with their support contained in $\{\leq (k-\Delta)\}$ and $\{>(k+\Delta)\}$, respectively, for any $k=\Delta,..,n+1-\Delta$. Then it holds that
\begin{equation}\label{ProdRel}
\langle  \Phi |AB|\Phi \rangle = \langle \Phi |A|\Phi\rangle  \langle \Phi |B|\Phi\rangle. 
\end{equation}
\end{Lemma}
{\em Proof of Lemma \ref{Prod}.} See Ref. \onlinecite{SOCO}. $\Box$ \\

As a result of Lemma \ref{Prod}, there is a useful factorization property for products of string order operators, stated as follows. 
\begin{Lemma}\label{Lemma:string_factorize}
Let $U_\Phi$ be a short-range entangled quantum circuit with entanglement range $\xi$  preserving the $\mathbb{Z}_2\times \mathbb{Z}_2$ symmetry. 
Let $\Pi_{i=1}^m \mathcal{K}_{\geq k_i}$ and $\Pi_{j=1}^n \mathcal{K}_{\geq l_j}$ be two products of string order operators, where $l_j-k_i> \xi$, $\forall 1\leq i\leq m, 1\leq  j\leq n$.
Then
\begin{eqnarray}
\bra{\Phi}\Pi_{i=1}^m \mathcal{K}_{\geq k_i}\cdot \Pi_{j=1}^n \mathcal{K}_{\geq l_j}\ket{\Phi}
=\bra{\Phi}\Pi_{i=1}^m \mathcal{K}_{\geq k_i}\ket{\Phi}
\bra{\Phi} \Pi_{j=1}^n \mathcal{K}_{\geq l_j}\ket{\Phi}.
\label{eq:factorize_string}
\end{eqnarray}
\end{Lemma}

{\em Proof of Lemma \ref{Lemma:string_factorize}.}
Define $\bar{k}$ to be the smallest even integer larger than $\text{max}\{k_i|1\leq i \leq m\}$, and $\underline{l}$ to be the largest odd integer smaller than  $\text{min}\{l_j|1\leq j\leq n\}$.
Restricting $\Pi_{i=1}^m \mathcal{K}_{\geq k_i}\cdot \Pi_{j=1}^n \mathcal{K}_{\geq l_j}$ to the sites within the interval $[\bar{k},\underline{l}]$, 
we have four possibilities, i.e., $IIII...II$, $XIXI...XI$, $IXIX...IX$ or $XXXX...XX$,
if (number of even values of $k_i$'s, number of odd values of $k_i$'sz) is (even, even), (odd, even), (even, odd), (odd, odd), respectively. 
Consider $g\cdot \Pi_{i=1}^m \mathcal{K}_{\geq k_i}\cdot \Pi_{j=1}^n \mathcal{K}_{\geq l_j}$
in which $g=1,g_0,g_1,g_0g_1$ for the (even, even), (odd, even), (even, odd) and (odd, odd) cases, respectively,
where $g_0=Z_1 X_2 X_4 \ldots  X_{N-1} Z_N$ and $g_1=X_1  X_3 \ldots X_{N-2} X_N$ are the generators of the $\mathbb{Z}_2\times\mathbb{Z}_2$ symmetry. 
Then it is easy to see that $\text{supp}(g \Pi_{i=1}^m \mathcal{K}_{\geq k_i}) \subset \{\leq \bar{k}\}$
and $\text{supp}(\Pi_{j=1}^n \mathcal{K}_{\geq l_j})\subset\{\geq \underline{l}\}$. 
Applying Lemma \ref{Prod}, we obtain 
\begin{eqnarray}
\bra{\Phi}\Pi_{i=1}^m \mathcal{K}_{\geq k_i}\cdot \Pi_{j=1}^n \mathcal{K}_{\geq l_j}\ket{\Phi}
=\bra{\Phi}g\Pi_{i=1}^m \mathcal{K}_{\geq k_i}\ket{\Phi}
\bra{\Phi} \Pi_{j=1}^n \mathcal{K}_{\geq l_j}\ket{\Phi},
\end{eqnarray}
which gives Eq. (\ref{eq:factorize_string}) by using $\bra{\Phi}g=\bra{\Phi}$ since $g$ is a symmetry.  $\Box$\\

\begin{Cor}\label{Cor:string_factorize}
Consider $t$ group of sites $\{k^{(i)}_j| 1 \leq j\leq n_i\}$ ($1\leq i\leq t$) separated by distances larger than $\xi$, i.e.,
$k^{(i)}_{j}-k^{(p)}_{q}>\xi$, $\forall i> p, 1 \leq j\leq n_i, 1 \leq q\leq n_p$. 
Then 
\begin{equation}\label{eq:repeat}
\bra{\Phi} \Pi_{i=1}^t (\Pi_{j=1}^{n_i} \mathcal{K}_{\geq k^{(i)}_j})\ket{\Phi} =\Pi_{i=1}^t\bra{\Phi} \Pi_{j=1}^{n_i} \mathcal{K}_{\geq k^{(i)}_j}\ket{\Phi}.
\end{equation}
\end{Cor}

{\em Proof of Corollary  \ref{Cor:string_factorize}.} It is straightforward to prove Eq. (\ref{eq:repeat}) by repeatedly applying Lemma \ref{Lemma:string_factorize}. $\Box$\\

Before closing this section, we note that 
$\expval*{\mathcal{K}_{\geq x}}$ is independent of $x$ and $\expval*{\mathcal{K}_{\geq x}\mathcal{K}_{\geq y}}=\expval*{\mathcal{K}_{|x-y|}}$ only depends on $|x-y|$
when $x$ and $y$ are far from the boundaries,
since the boundary effects decay exponentially in the bulk for a short-range entangled resource state.

\section{:  Product formula for logical operations}
\label{sec:product_formula}

To derive a formal expression for the evolution  of the logical qubit through a $m$-site rotation (region $\mathcal{R}$), let us consider the setting displayed in Fig.~\ref{Formal Setting}. The first $(p+1)$-sites in the chain are used for preparation of some arbitrary initial state, the next $d-1$ sites, measured in the symmetry respecting axis, act as a buffer, where both $p$ and $d$ are odd integers. Furthermore, the next $m\Delta$ sites (denoted as $\mathcal{R}$ in the figure) are used to perform the desired logical operation ($z$-rotation for our example) where $\Delta$ is always an even integer,
and the computational output is read out from the last site in the chain. 
\begin{center}
\begin{figure}[H]
    \centering
    \begin{tikzpicture}[scale=1] 
        \node[] at (-0.25, 0.5) {\phantom{i}};
        \node[] at (0, 1) {\phantom{i}};
        \filldraw[thick, draw = blue!80, fill = blue!20] (0, 0) -- (4, 0) -- (4, 1) -- (0, 1) -- cycle;
        \filldraw[thick, draw = red!80, fill = red!20] (7, 0) -- (11, 0) -- (11, 1) -- (7, 1) -- cycle;
        \node[below, blue] at (2, 0) {preparation};
        \node[below, red] at (9, 0) {$\mathcal{R}$};
        \node[below, red] at (9, -0.35) {$m$-site rotation};
        \node[] at (5.5, 0.2) {$\underbrace{\phantom{AAAAAAAA}}$};
        \node[below] at (5.5, 0) {buffer};
        \node[color=green!70!black, below] at (13.5, 0) {readout};
        \foreach \i in {0, 1, 2, ..., 12}{
	           \draw[thick] (\i + 0.5, 0.5) circle (0.2);
        };
        \filldraw[fill=green!70!black, draw=green!70!black] (13.5, 0.5) circle (0.2);
        \foreach \i in {0, 2, 3, 4, 6, 8, 10, 12}{
	           \draw[thick] (\i + 0.7, 0.5) -- (\i + 1.3, 0.5);
        };
        \foreach \i in {1, 5, 7, 9, 11} {
            \node[] at (\i + 1.03, 0.5) {$\cdots$};
        }
    \end{tikzpicture}  
    \caption{The spin chain of length $N$ as a computational resource: the spins through $-p$ to $0$ are used for preparation of an arbitrary initial state, the sites from $d$ to $d+m\Delta$ are used for performing a rotation about $z(x)$-axis and the last site $N$ is used for readout. The $d$ sites (chosen to be ``large") in the middle are used as a buffer between the initial state and the logical operation. This is only place in the analysis where short-range entanglement plays a role.}
    \label{Formal Setting}
\end{figure}
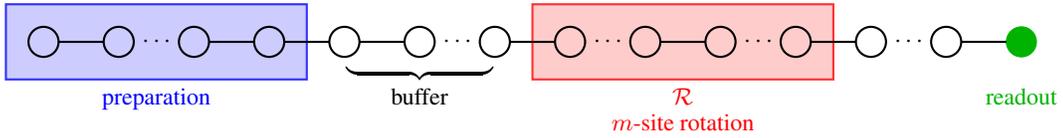
  \end{center}
As some background information, recall that the measured observables in the cluster phase are of the form
\begin{equation}\label{Observables}
  O_i(q_i,\gamma_i):=\cos\gamma_i X_i-(-1)^{q_i}\sin \gamma_i Y_i.  
\end{equation}
The sign flippers $q_i$'s can be related to measurement outcomes $s_i$ of the qubits in the chain in a linear fashion. Furthermore, the one bit $o$ of computational output is related to the measurement record (for odd $N$) as

$$o= \begin{cases}\sum_{i \text { odd }}^{N-1} s_i \bmod 2, & \text { if readout in } x \text {-basis, } \\ \sum_i^{N-1} s_i \bmod 2, & \text { if readout in  } y \text {-basis, } \\ \sum_{i \text { even }}^{N-1} s_i \bmod 2, & \text { if readout in } z \text {-basis. }\end{cases}$$

If the eigenstates of $O_i$ defined in Eq.~\ref{Observables} are labelled as $|\gamma_i,q_i,s_i\rangle$, then the global measurement basis is $\mathcal{B}=\{|\mathbf{s}\rangle,\mathbf{s}\in \mathbb{Z}_2^N\}$ 
$$|\mathbf{s}\rangle:=\bigotimes_{i=1}^{N}|\gamma_i,q_i(\mathbf{s}),s_i\rangle$$

The decorated observables that take into account the classical side-processing of MBQC can thus be simply defined as
\begin{equation}\label{DecoratedObservables}
 \tilde{O}_i=\sum_{\mathbf{s} \in \mathbb{Z}_2^N}(-1)^{s_i}|\mathbf{s}\rangle\langle\mathbf{s}|, \forall i, i<N   
\end{equation}

Finally, the non-local logical  Pauli observables whose measurement yields the computational output $o$ are defined as

\begin{equation}\label{PauliDefs}
    \overline{\mathcal{X}}:= X_N \bigotimes_{i \text{ odd}}^{N-1} \tilde{O_i},\quad  \overline{\mathcal{Y}}:= Y_N \bigotimes_{i }^{N-1} \tilde{O_i}, \quad \overline{\mathcal{Z}}:= Z_N \bigotimes_{i \text{ even}}^{N-1} \tilde{O_i}
\end{equation}

\subsection{Recursive Relation for arbitrary rotations}

Applying the above relation to the case where the all the sites (in the bulk) after $d+q\Delta$ are measured in the symmetry respecting basis,

\begin{align*}
  \overline{\mathcal{X}}_q&= \left(\bigotimes_{\substack{-p\leq i< d+q\Delta \\ \text{odd}}}  \tilde{O}_i (\gamma_i)\right)  \tilde{O}_{d+q\Delta}\left(\bigotimes_{\substack{d+q\Delta<k<N \\ \text{odd}}} X_k \right)X_N \\
  \overline{\mathcal{Y}}_q&= \left(\bigotimes_{-p\leq i< d+q\Delta}  \tilde{O}_i (\gamma_i)\right) \tilde{O}_{d+q\Delta}\left(\bigotimes_{d+q\Delta<k<N} X_k \right)Y_N \\
\end{align*}

When an additional symmetry breaking measurement is made on the $d+(q+1)\Delta$ th site, we have
\begin{align*}
    \overline{\mathcal{X}}_{q+1}&= \left(\bigotimes_{\substack{-p\leq i< d+(q+1)\Delta \\ \text{odd}}}\tilde{O}_i(\gamma_i) \right)  \tilde{O}_{d+(q+1)\Delta}\left(\bigotimes_{\substack{d+(q+1)\Delta<k_2<N \\ \text{odd}}} X_{k_2} \right)X_N 
\end{align*}

Now for the site $a\equiv d+(q+1)\Delta$:
\begin{align*}
    \tilde{O}_{a}= \cos\gamma_{a} X_a - \sin \gamma_a \left(\bigotimes_{\substack{-p\leq i<a \\ \text{even}}}  \tilde{O}_i (\gamma_i) \right)  Y_a
\end{align*}

\begin{enumerate}
    \item The term $\propto \cos \gamma_{d+(q+1)\Delta}$ in $\overline{\mathcal{X}}_{q+1}$ is merely $\overline{\mathcal{X}}_{q}$
    \item The term $\propto \sin \gamma_{d+(q+1)\Delta}$ in $\overline{\mathcal{X}}_{q+1}$ can be written as: $\mathcal{K}_{\geq d+( {q+1})\Delta}\overline{\mathcal{Y}}_q$
    
\end{enumerate}

Thus, overall we arrive at the recursion relation

\begin{equation*}
    \overline{\mathcal{X}}_{q+1}=\cos \gamma \overline{\mathcal{X}}_{q} -\sin \gamma \mathcal{K}_{\geq d+(q+1)\Delta}\overline{\mathcal{Y}}_{q}
\end{equation*}

Analogous analysis yields

\begin{equation*}
    \overline{\mathcal{Y}}_{q+1}=\sin \gamma \mathcal{K}_{\geq d+(q+1)\Delta}\overline{\mathcal{X}}_{q}+ \cos \gamma \overline{\mathcal{Y}}_{q}
\end{equation*}

and $\overline{\mathcal{Z}}_{q+1}=\overline{\mathcal{Z}}_{q}$ which is better represented in matrix form as

\begin{equation}
  \begin{pmatrix}
     \overline{\mathcal{X} }_{q+1}\\
     \overline{\mathcal{Y} }_{q+1}\\
     \overline{\mathcal{Z} }_{q+1}
  \end{pmatrix}  = M_{q+1} \begin{pmatrix}
     \overline{\mathcal{X} }_{q}\\
     \overline{\mathcal{Y} }_{q}\\
     \overline{\mathcal{Z} }_{q}
  \end{pmatrix}.
\end{equation}
 wherein $$M_{q+1}(\gamma):=\begin{pmatrix}
\cos\gamma & -\sin\gamma\;\mathcal{K}_{\geq d+(q+1)\Delta} & 0 \\ \sin\gamma\;\mathcal{K}_{\geq  d+(q+1)\Delta}  & \cos\gamma & 0 \\
0& 0 & 1
\end{pmatrix} $$
Since all the sites not belonging to the preparation and the set $\mathcal{R}$ are measured in the symmetry respecting basis, they correspond to a quantum wire operation where the logical qubit goes through the identity gate. Thus, applying the above relation recursively, we  arrive at the expectation values of the Pauli observables at the end of the entire evolution 

\begin{equation}
\displaystyle
    \begin{pmatrix}
\expval*{\overline{\mathcal{X}}} \\ 
\expval*{\overline{\mathcal{Y}}} \\
\expval*{\overline{\mathcal{Z}}} 
\end{pmatrix}= \Biggl<\left(\prod_{k\in \mathcal{R}}M_k  \right). \begin{pmatrix}
\overline{\mathcal{X}}_{0} \\ \overline{\mathcal{Y}}_{0} \\
\overline{\mathcal{Z}}_{0}
\end{pmatrix}\Biggr>
\end{equation}

The above expression can be further simplified using the fact the state is short-range entangled and $d$ is chosen to be much larger than the correlation length.

\begin{equation}\label{FactoredPaulis}
\displaystyle
   \boxed { \begin{pmatrix}
\expval*{\overline{\mathcal{X}}} \\ \expval*{\overline{\mathcal{Y}}} 
\\ \expval*{\overline{\mathcal{Z}}} 
\end{pmatrix}= \biggl<\left(\prod_{k\in \mathcal{R}}M_k  \right)\biggr>. \begin{pmatrix}
\expval*{\overline{\mathcal{X}}_0} \\ \expval*{\overline{\mathcal{Y}}_0} \\
\expval*{\overline{\mathcal{Z}}_0} 
\end{pmatrix}  }
\end{equation}

This comes about as follows.
\vspace{5pt}
The terms containing ${\overline{\mathcal{X}}_0}$ above can be expressed as $ \propto \expval*{\overline{\mathcal{X}}_0. S_{\geq d }} $ where $S_d$ is some (product of string order) operator with non-trivial support on only sites $\geq d$. Using the symmetry element $g_1$, 

\begin{equation*}
    \expval*{\overline{\mathcal{X}}_0. S_{\geq d }}
    = \biggl<{\left(\bigotimes_{i=-p}^0 \tilde{O}_i X_i\right). \left(S_{\geq d }\right)}\biggr>
    =\biggl<{\left(\bigotimes_{i=-p}^0 \tilde{O}_i X_i\right)}\biggr>. \expval*{S_{\geq d }}
    = \expval*{\overline{\mathcal{X}}_0}.\expval*{ S_{\geq d }}
\end{equation*}

wherein we have used the short-range entangled nature of our resource state in the second equality. Using the symmetry element $g_0.g_1$ we find a similar expression 

\begin{equation}
    \expval*{\overline{\mathcal{Y}}_0. S_{\geq d }}=\expval*{\overline{\mathcal{Y}}_0}.\expval*{ S_{\geq d }}
\end{equation}

Furthermore, since the $z$ component is unchanged under this evolution,  this proves Eq.~\ref{FactoredPaulis}.\qed

For rotations about the $x$-axis using even sites on the chain, the analysis still holds with the expression for $M_k$ replaced by

\begin{equation*}
M_k(\gamma):=\begin{pmatrix}
1 &0 &0 \\
0 &\cos\gamma & -\sin\gamma\;\mathcal{K}_{\geq d+1+ k\Delta}  \\ 0 & \sin\gamma\;\mathcal{K}_{\geq d+1+ k\Delta}  & \cos\gamma  
\end{pmatrix} . 
\end{equation*}

\section{:  Proof of Theorem 1}\label{PT1}

Theorem 1 in the main text asserts: ``Throughout the 1D cluster phase with $\mathbb{Z}_2 \times \mathbb{Z}_2$ symmetry, for all separations $\Delta>0$ between successive symmetry-breaking measurements, the logical error $\epsilon$ for an $x$- or $z$-rotation about an angle $\beta$ can be made arbitrarily small, by splitting the rotation into $m$ steps. The logical error is $\mathcal{D}_m = \beta_{\text{log}}^2 \,\kappa/m+ O(1/m^2) $, 
with
$$
\kappa = \left(1+2\sum_{j=1}^{m-1}f(j\Delta)\right)\left(\frac{1}{\expval*{\mathcal{K}_{\geq}}^2}-1\right)\;
$$
a constant depending on the string order parameter. Therein, $f(\Delta):= (\expval*{\mathcal{K}_{\Delta}}-{\expval*{\mathcal{K}_{\geq}}}^2 )/(1-{\expval*{\mathcal{K}_{\geq}}}^2)$."
In this section, we give a proof for Theorem 1. 

We first introduce a measure for the decoherence related to single qubit operations. 
Density matrices for a single qubit can be expanded in the basis $\{I,X,Y,Z\}$, where $I$ is the $2\times 2$ identity matrix. 
Since quantum operations are linear maps on the density matrices, a general quantum operation $\mathcal{M}$ for a single qubit can be represented by a $4\times 4$ matrix $M$ in the basis $\{I,X,Y,Z\}$ in the following way
\begin{eqnarray}
(\mathcal{M}(I),\mathcal{M}(X),\mathcal{M}(Y),\mathcal{M}(Z))=(I,X,Y,Z)M. 
\label{eq:M_M}
\end{eqnarray}
If we restrict ourselves to trace-preserving quantum operations, then $I$ is always mapped  to $I$.
%In such cases, the aforementioned $4\times 4$ matrix can be reduced to a $3\times 3$ one in the basis $\{X,Y,Z\}$. 

Suppose we want to realize a unitary SU$(2)$ rotation $\mathcal{R}$.
In this case, the matrix in Eq. (\ref{eq:M_M}) reduces to a $3\times 3$ one represented by $R$. 
It is not hard to verify that for a rotation around $\hat{n}$-axis by $\theta$,
the operation $\mathcal{R}$ and the matrix $R$ can be related by
\begin{eqnarray}
(\mathcal{R}(X),\mathcal{R}(Y),\mathcal{R}(Z))=(X,Y,Z)R,
\end{eqnarray}
in which $R=e^{i\theta\vec{J}\cdot \hat{n}}$,
and $\vec{J}=(J^x,J^y,J^z)$ are the $3\times 3$  matrices given by
$(J^\alpha)_{\beta\gamma}=-i\epsilon_{\alpha\beta\gamma}$, 
where $\epsilon_{\alpha\beta\gamma}$ is the totally anti-symmetric tensor and the indices $1,2,3$ of the rows and columns of the matrix are identified with $x,y,z$, respectively. 
For example, considering $R=R(\hat{z},\beta_{\text{log}})$ as a $z$-rotation by an angle $\beta_{\text{log}}$,
then we have
\begin{eqnarray}
R(\hat{z},\beta_{\text{log}})=\left(\begin{array}{ccc}
\cos(\beta_{\text{log}}) & -\sin(\beta_{\text{log}})&0\\
\sin(\beta_{\text{log}}) & \phantom{+}\cos(\beta_{\text{log}})&0\\
0&0&1
\end{array}\right).
\label{eq:Rz}
\end{eqnarray}

In practice, only a non-unitary quantum operation $\mathcal{M}_{\mathcal{R}}$ can be implemented which gives an approximated realization for $\mathcal{R}$.
To quantitatively evaluate how good $\mathcal{R}$ is approximated by $\mathcal{M}_{\mathcal{R}}$,
we introduce the following metric to characterize the distance between two quantum operations.

\begin{Def}\label{def:metric}
The distance $d(\mathcal{R},\mathcal{M})$ between two  single-qubit quantum operations $\mathcal{R}$ and $\mathcal{M}$ is defined as
\begin{eqnarray}
d(\mathcal{R},\mathcal{M})= \sqrt{2\text{Tr}[(M-R)^\dagger (M-R)] },
\end{eqnarray}
in which $R$ and $M$ are the corresponding $4\times 4$ matrices determined by Eq. (\ref{eq:M_M}).
\end{Def}
  
Using Definition \ref{def:metric}, 
the decoherence $\mathcal{D}$ can be defined as
\begin{eqnarray}
\mathcal{D}=d(\mathcal{M}_\mathcal{R},\mathcal{R}).
\end{eqnarray}
For example, considering $R=R(\hat{z},\beta_{\text{log}})$ and splitting the angle $\beta_{\text{log}}$ into $m$ pieces, the decoherence function $\mathcal{D}_m$ can be evaluated as
\begin{eqnarray}
\mathcal{D}_m=\sqrt{2\sum_{ij}(M_{ij}-R_{ij})^2},
\label{eq:D_m}
\end{eqnarray}
in which $R$ is given in Eq. (\ref{eq:Rz})
and $M$ is 
\begin{eqnarray}\label{eq:M}
M=\bigl<{\Pi_{k=0}^{m-1} M_k(\gamma)}\bigr>.
\end{eqnarray}
In Eq. (\ref{eq:M}), the expression of $M_k$ is given by
\begin{eqnarray}
M_k=\left(\begin{array}{ccc}
\cos(\gamma_k) & -\sin(\gamma_k) \mathcal{K}_{\geq k\Delta}&0\\
\sin(\gamma_k) \mathcal{K}_{\geq k\Delta}&\cos(\gamma_k)&0\\
0&0&1
\end{array}\right),
\end{eqnarray}
in which $\gamma_k=\beta_{\text{log}}/(m\expval*{\mathcal{K}_\geq})$,
and the reference point (i.e., the origin of the spatial coodinate) is shifted by a distance $d$ compared with Sec. \ref{sec:product_formula}.
It is worth emphasising that to approximately implement a rotation by an angle $\beta_{\text{log}}$ in the SPT phase, the angle on each rotation site is not $\beta_{\text{log}}/m$, but scaled up by a factor of $1/\expval*{\mathcal{K}_\geq}$,
which is proved rigorously in Corollary \ref{Cor:beta}.

%We also note that for the scenario depicted in Fig. \ref{Formal Setting},
%since the $m$ sites used for implementing a rotation are far away from the boundaries, 
% $\expval*{\mathcal{K}_{\geq j}}$ is independent of $j$ (denoted as $\expval*{\mathcal{K}_\geq}$)
%and $\expval*{\mathcal{K}_{\geq i} \mathcal{K}_{\geq j}}$ only depends on $|i-j|$ (denoted as $\expval*{\mathcal{K}_{|i-j|}}$),
%when $i,j$ both belong to the group of $m$ sites. 

Next we derive an exact formula for the decoherence function  $\mathcal{D}_m$ as summarized in the following Lemma.

\begin{Lemma} 
\label{Lemma:expression_Dm}
The decoherence function $\mathcal{D}_m$ for $z$-rotations defined in Eq. (\ref{eq:D_m}) can be expressed as
\begin{eqnarray}
\mathcal{D}_m=2 |\bigl<{e^{i\gamma\sum_{k=0}^{m-1} (\mathcal{K}_{\geq k\Delta}-\expval*{\mathcal{K}_\geq}) }}\bigr> -1|,
\label{eq:D_m_exact}
\end{eqnarray}
in which $\expval*{\mathcal{K}_\geq}= \expval*{\mathcal{K}_{\geq j\Delta}}$ ($ 0 \leq j\leq m-1$) and $\gamma=\beta_{\text{log}}/(m\expval*{\mathcal{K}_\geq})$. 
\end{Lemma}

{\em Proof of Lemma \ref{Lemma:expression_Dm}.}
For $z$-rotations, it is enough to consider the upper-left $2\times 2$ block of $M$. 
It is straightforward to see that $M_k$ can be diagonalized as
\begin{eqnarray}
M_k= U \left(\begin{array}{cc}
e^{-i\gamma \mathcal{K}_{\geq k\Delta} } & 0\\
0& e^{i\gamma \mathcal{K}_{\geq k\Delta} }
\end{array}\right) U^{-1},
\end{eqnarray}
in which  $\cos\gamma  + i \sin\gamma \;\mathcal{K}_{\geq k\Delta}= e^{i \gamma \mathcal{K}_{\geq k\Delta}}$ is used, 
$\gamma$ is given by $\gamma=\beta_{\text{log}}/(m\expval*{\mathcal{K}_\geq})$, 
and
\begin{eqnarray}
U=\frac{1}{\sqrt{2}}\left(\begin{array}{cc}
1 & 1\\
i &-i
\end{array}
\right).
\end{eqnarray}
Using the fact that $M_k$'s mutually commute with each other for $0\leq k \leq m-1$, we obtain
\begin{equation*}
M= \biggl<{U
\left(\begin{array}{cc}
e^{-i\gamma\sum_{k=0}^{m-1} \mathcal{K}_{\geq k\Delta} } & 0\\
0& e^{i\gamma \sum_{k=0}^{m-1}\mathcal{K}_{\geq k\Delta} }
\end{array}\right)
U^{-1}}\biggr>,
\end{equation*}
which gives
\begingroup
\renewcommand*{\arraystretch}{2.5}
\begin{equation}
M=\begin{pmatrix}
\biggl<{\cos(\gamma\sum_{k=0}^{m-1} \mathcal{K}_{\geq k\Delta})}\biggr> & -\biggl<{\sin(\gamma\sum_{k=0}^{m-1} \mathcal{K}_{\geq k\Delta})}\biggr> \\ 
\biggl<{\sin(\gamma\sum_{k=0}^{m-1} \mathcal{K}_{\geq k\Delta})}\biggr>  & \phantom{-}\biggl<{\cos(\gamma\sum_{k=0}^{m-1} \mathcal{K}_{\geq k\Delta})}\biggr>

\end{pmatrix}
\end{equation}
\endgroup

As a result of Definition \ref{def:metric}, we obtain
\begin{eqnarray}
\label{eq:Dm_1}
\mathcal{D}_m=2\sqrt{\left[\biggl<{\cos(\gamma\sum_{k=0}^{m-1} \mathcal{K}_{\geq k\Delta})}\biggr>-\cos(\beta_{\text{log}})\right]^2+\left[\biggl<{\sin(\gamma\sum_{k=0}^{m-1} \mathcal{K}_{\geq k\Delta})}\biggr>-\sin(\beta_{\text{log}})\right]^2},
\end{eqnarray}
which can be simplified as
\begin{eqnarray}
\mathcal{D}_m=2\sqrt{ G_m(G_m)^*}, 
\end{eqnarray}
where $G_m$ is defined as
\begin{eqnarray}
\label{eq:def_Gm}
G_m=\bigl<{e^{i\gamma\sum_{k=0}^{m-1} (\mathcal{K}_{\geq k\Delta}-\expval*{\mathcal{K}_\geq}) }}\bigr> -1.
\end{eqnarray} 
This proves Eq. (\ref{eq:D_m_exact}).  $\Box$ \\

We perform a formal series expansion of $G_m$ in Eq. (\ref{eq:def_Gm}) over $\gamma$ 
\begin{eqnarray}
G_m=\sum_{n=0}^\infty \gamma^n G_m^{(n)},
\label{eq:Gm_expansion}
\end{eqnarray}
in which $G_m^{(n)}$ is given by
\begin{eqnarray}
G_m^{(n)}=\frac{i^n}{n!}\sum_{k_1,k_2,...,k_n=0}^{m-1} \expval*{
(\mathcal{K}_{\geq k_1\Delta}-\expval*{\mathcal{K}_{\geq}})(\mathcal{K}_{\geq k_2\Delta}-\expval*{\mathcal{K}_{\geq}})...(\mathcal{K}_{\geq k_n\Delta}-\expval*{\mathcal{K}_{\geq}})
}.
\label{eq:def_Gm_n}
\end{eqnarray}
Plugging in $\gamma=\beta_{\text{log}}/(m\expval*{\mathcal{K}_\geq})$, $G_m$ can be rearranged in terms of a series expansion over $1/m$. 
Since $m$-dependence appears in both the factor $\gamma^n$ as well as the summation over $k_i$  ($1\leq i\leq n$),
the expansion over $1/m$ is not straightforward. 
However,  we will show that there is a simpler structure in the lowest order terms, as stated in the following lemma.

\begin{Lemma} 
\label{Lemma:1_over_m}
$G_m^{(n)}$ defined in Eq. (\ref{eq:def_Gm_n}) can be written as a power series of $1/m$, satisfying:
\begin{enumerate}
\item the $O(1)$ term vanishes in $G_m^{(n)}$ for all $n$, 
\item the $O(1/m)$ term vanishes in $G_m^{(n)}$ for $n\geq 3$. 
\end{enumerate}
\end{Lemma}

{\em Proof of Lemma \ref{Lemma:1_over_m}.}
\begin{enumerate}
\item 
Define $E_i(k_1,...,k_i) $ as
\begin{eqnarray}
E_i(k_1,...,k_i) =  (\mathcal{K}_{\geq k_1\Delta}-\expval*{\mathcal{K}_\geq})(\mathcal{K}_{\geq k_2\Delta}-\expval*{\mathcal{K}_\geq})...(\mathcal{K}_{\geq k_i\Delta}-\expval*{\mathcal{K}_\geq}).
\label{eq:def_E}
\end{eqnarray}
Then $G_m^{(n)}$ can be written as
\begin{eqnarray}
G_m^{(n)}=\frac{i^n}{n!}\sum_{k_1,k_2,...,k_n=0}^{m-1} \expval*{E_n(k_1,...,k_n)}.
\label{eq:G_m_n_E}
\end{eqnarray}

Next we introduce the notation $\sum_{(j_1,...,j_t)}$ with $j_1\leq j_2\leq ...\leq j_t$ and $\sum_{s=1}^t j_s=n$, defined  as
\begin{eqnarray}
\sum_{(j_1,...,j_t)}=\sum^\prime_P~\sum_{\{k_{P(1)}\sim k_{P(2)}\sim ...\sim k_{P(j_1)}\},~\{k_{P(j_1+1)}\sim k_{P(j_1+2)}\sim ...\sim k_{P(j_1+j_2)}\},...,\{k_{P(j_{1}+...+j_{t-1}+1)}\sim ...\sim k_{P(n)}\}},
\label{eq:sum_j1_to_jt}
\end{eqnarray}
in which the $k_j$'s connected by ``$\sim$" are mutually within a distance of $\xi$ (where $\xi$ is defined in Definition \ref{def:SRE_states}),
i.e., $|k_p-k_q|\leq \xi$ if $k_p$ and $k_q$ are within the same group of sites,
whereas different groups of $k_j$'s that are not connected by ``$\sim$" are separated by distances larger than $\xi$,
i.e., $|k_p-k_q|> \xi$ if $k_p$ and $k_q$ are from different groups. 
The summation $\sum^\prime_P$ is over all distinct divisions of the set $\{1,2,...,n\}$ into $t$ subsets $\{P(1),P(2),...,P(j_1)\}$, $\{P(j_1+1),P(j_1+2),...,P(j_1+j_2)\}$, ..., $\{P(j_1+j_2+...+j_{t-1}+1),P(j_1+j_2+...+j_{t-1}+2),...,n\}$, with number of elements equal to $j_1,j_2,...,j_t$, respectively. 
The summation over $k_j$'s ($1\leq j \leq n$) in Eq. (\ref{eq:G_m_n_E}) can be decomposed as
\begin{eqnarray}
\sum_{k_1,...,k_{n}=0}^{m-1}=\sum_{j_1+...+j_t=n,1\leq j_1\leq...\leq j_t} ~\sum_{(j_1,...,j_t)},
\label{eq:sum_decompose}
\end{eqnarray}
in which $\sum_{j_1+...+j_t=n,1\leq j_1\leq...\leq j_t}$ denotes the summation over all possible partitions of $n$.

Since $(\mathcal{K}_{\geq k\Delta})^2=I$ where $I$ is the identity operator, $\mathcal{K}_{\geq k\Delta}$ is a bounded operator and its operator norm satisfies $||\mathcal{K}_{\geq k\Delta}||\leq 1$. 
As a result,
\begin{eqnarray}
||E_n(k_1,...,k_n)|| &\leq& \Pi_{j=1}^n ||\mathcal{K}_{\geq k_j\Delta}-\expval*{\mathcal{K}_\geq}||\nonumber\\
&\leq &  \Pi_{j=1}^n (||\mathcal{K}_{\geq k_j\Delta}||+|\expval*{\mathcal{K}_\geq}|)\nonumber\\
&\leq & (1+|\expval*{\mathcal{K}_\geq}|)^n,
\end{eqnarray}
which shows that $E_n(k_1,...,k_n)$ is also a bounded operator. 

For fixed $P$ in Eq. (\ref{eq:sum_j1_to_jt}), we choose one position variable from each group, denoted as $k^{(l)}$, $1\leq l \leq t$. 
The ranges of summations over $k^{(l)}$'s are all on order of $m$.
The summations over the remaining variables in the $l$'th group are restricted within a distance of $\xi$ from $k_l$.  
Hence 
\begin{flalign}\label{eq:order_m_t}
&|\sum_{\{k_{P(1)}\sim k_{P(2)}\sim ...\sim k_{P(j_1)}\},...,\{k_{P(j_{1}+...+j_{t-1}+1)}\sim ...\sim k_{P(n)}\}} \expval*{E_n(k_1,...,k_n)}|\nonumber\\
&\leq \sum_{\{k_{P(1)}\sim k_{P(2)}\sim ...\sim k_{P(j_1)}\},...,\{k_{P(j_{1}+...+j_{t-1}+1)}\sim ...\sim k_{P(n)}\}} |\expval*{E_n(k_1,...,k_n)}|\nonumber\\
&< (1+|\expval*{\mathcal{K}_\geq}|)^n\Pi_{l=1}^t (m \xi^{j_l-1}) \nonumber\\
&= (1+|\expval*{\mathcal{K}_\geq}|)^n m^t \xi^{n-t}. 
\end{flalign}

From Eq. (\ref{eq:order_m_t}), we see that $\sum_{(j_1,...,j_t)}$ is of order $O(m^t)$, hence the $O(1)$ term of $G_m^{(n)}$ in Eq. (\ref{eq:G_m_n_E}) comes from the $\sum_{(1,...,1)}$ term where $j_1=...=j_{n}=1$.
In this case, there is only one division, i.e., the summation $\sum_P^\prime$ in Eq. (\ref{eq:sum_j1_to_jt}) contains only one term.  
Using the factorization property in Corollary \ref{Cor:string_factorize}, 
we obtain
\begin{eqnarray}
G_m^{(n)}& =& \sum_{(1,1,...,1)}\Pi_{j=1}^{n} (\expval*{\mathcal{K}_{\geq k_j\Delta}} -\expval*{\mathcal{K}_\geq}) +O(1/m) \nonumber\\
&=&O(1/m),
\end{eqnarray}
in which $\expval*{\mathcal{K}_{\geq k_j\Delta}} =\expval*{\mathcal{K}_\geq}$ is used for long enough chain. 
Hence there is no $O(1)$ contribution in $G_m^{(n)}$ for any value of $n\geq 1$.

\item 
In Eq. (\ref{eq:sum_decompose}), 
the $O(1/m)$ contribution to $G_m^{(n)}$ originates from $\sum_{(1,...,1,2)}$ where $j_1=...=j_{n-2}=1$, $j_{n-1}=2$.
Suppose the two $k_j$'s whose separation is less than $\xi$ are $k_p$ and $k_q$,
then according to Cor. \ref{Cor:string_factorize}, 
$\expval*{E_n(k_1,...,k_n)}$ factorizes as 
\begin{eqnarray}
\expval*{E_n(k_1,...,k_n)}=\expval*{(\mathcal{K}_{\geq k_{p}\Delta}-\expval*{\mathcal{K}_\geq})(\mathcal{K}_{\geq k_{q}\Delta}-\expval*{\mathcal{K}_\geq}) }\Pi_{1\leq j\leq n,j\neq p,q}(\expval*{\mathcal{K}_{\geq k_{j}\Delta}}-\expval*{\mathcal{K}_\geq}).
\end{eqnarray}
Clearly, after performing the summation $\sum_{\{k_{P(1)}\},...,\{k_{P(n-2)}\},\{k_p,k_q\}}$, different choices of $p,q$ all contribute the same values. 
Since there are $C_n^2=n(n-1)/2$ ways to choose the pair $\{k,q\}$, 
we obtain 
\begin{eqnarray}
G_m^{(n)}=\frac{n(n-1)}{2}\expval*{(\mathcal{K}_{\geq k_{p}\Delta}-\expval*{\mathcal{K}_\geq})(\mathcal{K}_{\geq k_{q}\Delta}-\expval*{\mathcal{K}_\geq} )}\Pi_{1\leq j\leq n,j\neq p,q}(\expval*{\mathcal{K}_{\geq k_{j}\Delta}}-\expval*{\mathcal{K}_\geq})+O(1/m^2).
\end{eqnarray}

When $n\geq 3$, there is at least one factor in $\Pi_{1\leq j\leq n,j\neq p,q}(\expval*{\mathcal{K}_{\geq k_{j}\Delta}}-\expval*{\mathcal{K}_\geq})$. 
Using $\expval*{\mathcal{K}_{\geq k_{j}\Delta}}=\expval*{\mathcal{K}_\geq}$ for long enough chain, 
it is clear that $G_m^{(n)}=O(1/m^2)$ when $n\geq 3$. 
\end{enumerate}

This completes the proof of Lemma \ref{Lemma:1_over_m}. $\Box$ \\

\begin{Cor}\label{Cor:beta}
In the large $m$ limit, $M$ in Eq. (\ref{eq:M}) is approximately a $z$-rotation by an angle $\beta_{\text{log}}$. 
\end{Cor} 

{\em Proof of Corollary \ref{Cor:beta}.}
According to point 1 in Lemma \ref{Lemma:1_over_m}, the difference between $M$ and the matrix for $z$-rotation by an angle $\beta_{\text{log}}$ is on order of $1/m$.
Hence in the large $m$ limit, they are approximately equal. $\Box$\\

Now we are prepared to prove Theorem 1 in the main text.\\

{\em Proof of Theorem 1.}
In Eq. (\ref{eq:Gm_expansion}), $G_m^{(0)}$ is clearly zero.

It is straightforward to see that $G_m^{(1)}$ also vanishes since $\expval*{\mathcal{K}_{\geq k\Delta}}=\expval*{\mathcal{K}_{\geq}}$.

Next we calculate $G_m^{(2)}$. 
Plugging $n=2$ in Eq. (\ref{eq:def_Gm_n}), we obtain
\begin{eqnarray}
G_m^{(2)}&=&-\frac{\gamma^2}{2} \sum_{k_1,k_2=0}^{m-1} [\expval*{\mathcal{K}_{\geq k_1\Delta}\mathcal{K}_{\geq k_2\Delta}}-\expval*{\mathcal{K}_\geq }^2]\nonumber\\
&=& -\frac{\gamma^2}{2} [\sum_{k_1=k_2=0}^{m-1} \expval*{\mathcal{K}_{\geq k_1\Delta}\mathcal{K}_{\geq k_2\Delta}}+ \sum_{k_1\neq k_2=0}^{m-1} \expval*{\mathcal{K}_{\geq k_1\Delta}\mathcal{K}_{\geq k_2\Delta}} - m^2 \expval*{\mathcal{K}_\geq}^2]\nonumber\\
%&=& \gamma^2 [m(1-\expval{K_\geq}^2)+ \sum_{k_1\neq k_2=1}^m (\expval*{\mathcal{K}_{\geq k_1\Delta}\mathcal{K}_{\geq k_2\Delta}} - \expval{\mathcal{K}_\geq}^2)]\nonumber\\
&=&-\frac{\gamma^2}{2}(1-\expval{\mathcal{K}_\geq}^2) [m+ \sum_{k_1\neq k_2} f\big((k_1-k_2)\Delta\big)],
\end{eqnarray}
in which $(\mathcal{K}_{\geq k})^2=I$ is used, and $f(x)$ is defined as 
\begin{equation}\label{eq:def_f}
f(x):= (\expval*{\mathcal{K}{(x)}}-{\expval*{\mathcal{K}_{\geq}}}^2 )/(1-{\expval*{\mathcal{K}_{\geq}}}^2),
\end{equation}
where $\expval*{\mathcal{K}_x} =  \expval*{\mathcal{K}_{\geq k\pm x}\mathcal{K}_{\geq k}}$ in the long chain limit. 
Since 
\begin{eqnarray}
\sum_{k_1\neq k_2} f\big((k_1-k_2)\Delta\big)&=& \sum_{k_1\neq k_2} f\big((k_1-k_2)\Delta\big)\nonumber \\
&=& 2\sum_{k_1> k_2} f\big((k_1-k_2)\Delta\big)\nonumber \\
%&=& 2\sum_k\sum_{k_1>k} f\big(k\Delta\big)\nonumber \\
&=& 2\sum_{k=1}^m (m-k) f\big(k\Delta\big), 
\end{eqnarray}
we arrive at 
\begin{equation}
  {\displaystyle  G_m^{(2)} = -\frac{\gamma^2}{2} \; \left(1-\expval*{\mathcal{K}_{\geq}}^2\right)\;\left (m+2\sum_{j=1}^{m-1}(m-j)\,f(j\Delta)\right) }.
\end{equation}

To see the explicit dependence on $m$, it is instructive to write $G_m^{(2)}$ as
\begin{equation} 
    {\displaystyle  G_m^{(2)} =- \frac{\beta_{\text{log}}^2}{2} \; \left(\frac{1}{\expval*{\mathcal{K}_{\geq}}^2}-1\right)\; \left(\frac{1}{m}
    (1+2\sum\limits_{j=1}^{m-1}f(j\Delta))-\frac{2}{m^2}\sum\limits_{j=1}^{m-1} j. f(j\Delta)\right)}.
\end{equation}      
Keeping only the $O(1/m)$ term, we obtain 
\begin{equation} \label{LPGen}
    {\displaystyle  G_m^{(2)} =- \frac{\beta_{\text{log}}^2}{2} \; \left(\frac{1}{\expval*{\mathcal{K}_{\geq}}^2}-1\right)\; F(m)} +O\left(\frac{1}{m^2}\right),
\end{equation}      
in which
\begin{equation}
    F(m)=\frac{1}{m}\left(1+2\sum\limits_{j=1}^{m-1}f(j\Delta)\right)\quad .
\end{equation}

For $G_m^{(n)}$  ($n\geq 3$) terms in Eq. (\ref{eq:def_Gm_n}), 
according to Lemma \ref{Lemma:1_over_m}, they all decay with or faster than $1/m^2$. 
Therefore, the $G_m^{(2)}$ term contains all the $O(1/m)$ contribution to $\mathcal{D}_m=2|G_m|$.
As a result, $\mathcal{D}_m$ can be expanded as
\begin{equation}
\mathcal{D}_m = \kappa \beta_{\text{log}}^2/m+O(1/m^2), 
\end{equation}
in which $\kappa$ given by 
\begin{equation}
\kappa = \left(1+2\sum_{j=1}^{m-1}f(j\Delta)\right)\left(\frac{1}{\expval*{\mathcal{K}_{\geq}}^2}-1\right)
\end{equation}
is a constant depending on the string order parameter, where $f(x)$ is defined in Eq. (\ref{eq:def_f}). The $x$-rotation analysis is completely identical with a different string order parameter. 
 $\qed$

\section{:  Proof of Theorem 2}\label{PT2}

In the main text, Theorem 2 states: `` In MBQC on any state in the $\mathbb{Z}_2 \times \mathbb{Z}_2$ 1D cluster phase for which the string order parameter ${\cal{K}}(d)$ is a convex function of the distance $d$, the densest packing of symmetry-breaking measurements, $\Delta = 2$, is the most computationally efficient."
In this section, we prove this statement. 

Before starting the proof, we note an interesting Lemma, which, though not used in the proof of Theorem 2, sheds light on the properties of the string order parameters.

\begin{Lemma} 
\label{Lemma:decrease}
If $f(\Delta)$ is convex down, then it is a decreasing function for $\Delta\geq 0$.
\end{Lemma}

{\em Proof of Lemma \ref{Lemma:decrease}.}
Let $h(x)=f^\prime(x)$. 
Then by assumption, $h^\prime(x)\geq 0$ for all $x>0$. 
Furthermore, it is known that the string order parameter $f(x)$ decays to zero exponentially at large values of $x$,
hence $h(x)\rightarrow 0$ when $x\rightarrow \infty$.
%On the other hand, since $f(x)$ is bounded by $1$ from above and $f(0)=1$, we know that $h(0)<0$.
Suppose there exists $x_0\in (0,\infty)$ such that $h(x_0)$ is positive. 
On the other hand, we can choose sufficiently large $x_1>x_0$ such that $h(x_1)<h(x_0)$.
This clearly contradicts with the assumption that $h(x)$ is a non-decreasing function.
Hence $h(x)$ must be non-positive for all values of $x>0$. $\Box$ \\

A useful Lemma about convex down functions is the following. 

\begin{Lemma} 
\label{Lemma:convex_positive}
Let $h(x)$ be a convex down continuous function defined on interval $[a,b]$,
and $L(x)$ be the linear function on $[a,b]$ satisfying $L(a)=h(a)$, $L(b)=h(b)$.
Then $\int_{a}^b dx [L(x)-h(x)]\geq 0$.
\end{Lemma}

{\em Proof of Lemma \ref{Lemma:convex_positive}.}
 For $t\in[0,1]$, we have $x=tb+(1-t)a\in[a,b]$.
The convex-down property of $h(x)$ implies 
\begin{eqnarray}
h(x)\leq th(b)+(1-t) h(a).
\label{eq:h_leq}
\end{eqnarray}
Plugging  $t=\frac{x-a}{b-a}$ in Eq. (\ref{eq:h_leq}), we obtain
\begin{eqnarray}
h(x)\leq \frac{x-a}{b-a}h(b)+\frac{b-x}{b-a}h(a)=L(x).
\end{eqnarray}
Thus it is clear that $\int_{a}^b dx [L(x)-h(x)]\geq 0$ holds. $\Box$ \\

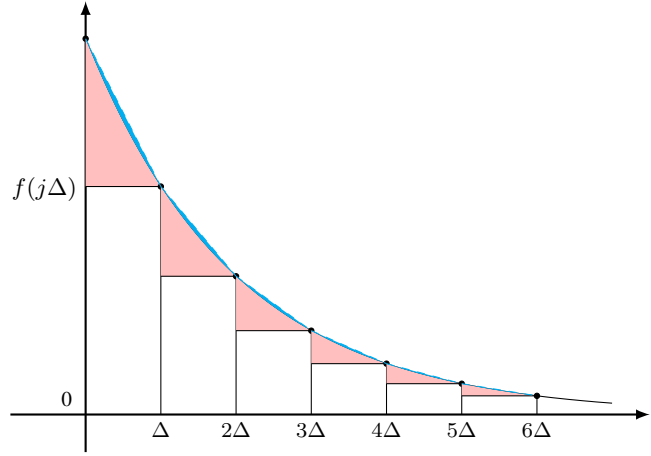
\begin{figure}
        \centering
        \begin{tikzpicture}[scale=1]
        \draw[thick, -latex] (0, -0.5) -- (0, 5.5);
        \draw[thick, -latex] (-1, 0) -- (7.5, 0);
        \draw[scale=1, domain=0:7, smooth, samples=100, variable=\x]  plot ({\x}, {5*exp(-0.5*\x)}); 
        \filldraw (0,5) circle (1pt);
        \node[above] at (-0.25, 0) {\footnotesize 0};
        \node[left] at (0, 3) {$f(j\Delta)$};
        \foreach \i in {1, 2, ..., 6} {
            \fill [pink, domain=\i-1:\i, variable=\x]
            (\i-1, {5*exp(-0.5*(\i-1))})
            -- plot ({\x}, {5*exp(-0.5*\x)})
            --  (\i-1, {5*exp(-0.5*(\i))})
            -- cycle;
            \fill [cyan, domain=\i-1:\i, variable=\x]
            (\i-1, {5*exp(-0.5*(\i-1))})
            -- plot ({\x}, {5*exp(-0.5*\x)})
            --  cycle;
            \draw[] (\i, 0) -- (\i, {5*exp(-0.5*\i)}) -- (\i-1, {5*exp(-0.5*\i)});
            \filldraw (\i,{5*exp(-0.5*\i)}) circle (1pt);
            \draw[dashed, cyan] (\i-1, {5*exp(-0.5*(\i-1))}) -- (\i, {5*exp(-0.5*\i)});
        }
        \node[below] at (1, 0) {\footnotesize$\Delta$};
        \foreach \x in {2, ..., 6} {
            \node[below] at (\x, 0) {\footnotesize $\x \Delta$};
        }
    \end{tikzpicture}
        \caption{Replacing the sum with a integral with error terms.The blue colored region represents the sum of the error terms that contain second or higher orders of $\Delta$. Let us call this sum over all the moon-edges $M_f(\Delta)$.}
        \label{SumIntegral}
    \end{figure}

Next we start proving Theorem 2. \\

{\em Proof of Theorem 2.} $m$ and $\Delta$ are dependent parameters via the constraint $m\Delta=n$ where $n$ is the constant number of physical qubits used to implement the $z$(or $x$)-rotation. Thus, we can always equivalently express
\begin{equation} \label{LPDelta}
    {\displaystyle  \mathcal{D}_m (\Delta) = \beta_{\text{log}}^2 \; \left(\frac{1}{\expval*{\mathcal{K}_{\geq}}^2}-1\right)\; F(\Delta)} +O(n^{-2}),
\end{equation} 
 where
\begin{eqnarray}
F(\Delta) = \frac{\Delta}{n}\left(1+ 2\sum\limits_{j\Delta=\Delta}^{n-\Delta} \, f(j\Delta) \right) = \frac{\Delta}{n}\left(1+ 2\sum\limits_{j\Delta=\Delta}^{\infty} \, f(j\Delta) \right),
\label{eq:decoherence_approx}
\end{eqnarray}
in which $n-\Delta$ is replaced by $\infty$ since $\sum_{j=n/\Delta}^\infty f(j\Delta)$ is exponentially small in $n$ in the long chain limit
and the difference is negligible when $n$ is much larger than the correlation length. 

Next, to separate the $\Delta$-independent term from the dependent ones in $F(\Delta)$, we rewrite the summation in the following way (see Fig.~\ref{SumIntegral})
\begin{eqnarray}
 \Delta \sum\limits_{j\Delta=\Delta}^{\infty} \, f(j\Delta)&=&\sum \text{area of white rectangles}\nonumber\\
 &=&\int_{0}^{\infty} f(x)dx-\sum \text{red regions}\nonumber\\
 &=&\int_{0}^{\infty} f(x)dx-\sum_{j=0}^{\infty} \frac{\Delta}{2}[f(j\Delta)-f((j+1)\Delta)]+ \sum \text{ moon-edge area},\nonumber\\
 &=&\int_{0}^{\infty} f(x)dx+\frac{\Delta}{2}[f(\infty)-f(0)]+ \sum \text{ moon-edge area},\nonumber\\
 &=&\int_{0}^{\infty} f(x)dx-\frac{\Delta}{2}+ \sum \text{ moon-edge area}.
\end{eqnarray}
in which ``moon-edge area" is used to denote the blue regions in Fig. \ref{SumIntegral}, and $f(\infty)=0$ and $f(0)=1$ are used.

To set up some notation, let us define the following quantities,
\begin{align}
    \mathcal{A}_f(i\Delta,j\Delta)&= \text{Area of the moon-edge between $i\Delta$ and $j\Delta$ } \\
    M_f(\Delta)&= \sum_{i=0}^{\infty} \mathcal{A}_f(i\Delta,(i+1)\Delta).
\end{align}
Then we obtain the following simplified expression of $F(\Delta)$
\begin{equation}
F(\Delta) = \frac{2}{n} \left(\int_{0}^{\infty} f(x)dx +  M_f(\Delta) \right).
\end{equation} 

It is graphically clear that $\mathcal{A}_f(i\Delta,(i+1)\Delta)\geq 0$,
which can be seen in a rigorous way.
Consider $h(x)=f(x)$, and $a=i\Delta$, $b=(i+1)\Delta$.
Notice that geometrically, the integral $\int_{i\Delta}^{(i+1)\Delta} dx [L(x)-f(x)]$ is exactly the moon-edge area $\mathcal{A}_f(i\Delta,(i+1)\Delta)$.
Hence $\mathcal{A}_f(i\Delta,(i+1)\Delta)$ must always be non-negative according to Lemma \ref{Lemma:convex_positive}.

\begin{figure}
        \centering
        \begin{tikzpicture}[scale=1]
        \draw[thick, -latex] (0, -0.5) -- (0, 5.5);
        \draw[thick, -latex] (-1, 0) -- (4.5, 0);
        \draw[scale=1, domain=0:4, smooth, samples=100, variable=\x]  plot ({\x}, {5*exp(-0.5*\x)}); 
        \filldraw (0,5) circle (1pt);
        \node[above] at (-0.25, 0) {\footnotesize 0};
        \node[left] at (0, 3) {$f(j\Delta)$};
        \foreach \i in {1, 2} {
            
            \fill [cyan, domain=\i-1:\i, variable=\x]
            (\i-1, {5*exp(-0.5*(\i-1))})
            -- plot ({\x}, {5*exp(-0.5*\x)})
            --  cycle;
            \draw[] (\i, 0) -- (\i, {5*exp(-0.5*\i)}) -- (\i-1, {5*exp(-0.5*\i)});
            \filldraw (\i,{5*exp(-0.5*\i)}) circle (1pt);
            \draw[dashed, cyan] (\i-1, {5*exp(-0.5*(\i-1))}) -- (\i, {5*exp(-0.5*\i)});    
        }
        \foreach \i in {2} {
            
            \fill [orange,opacity=0.2, domain=\i-2:\i, variable=\x]
            (\i-2, {5*exp(-0.5*(\i-2))})
            -- plot ({\x}, {5*exp(-0.5*\x)})
            --  cycle; 
        }

        \draw[dashed, orange] (0, {5}) -- (2, {5*exp(-0.5*2)});
            
        \node[below] at (1, 0) {\footnotesize$\Delta$};
        \foreach \x in {2} {
            \node[below] at (\x, 0) {\footnotesize $\x \Delta$};
        }
    \end{tikzpicture}
        \caption{Comparison of moon-edge areas: The blue coloured moon-edges denote the area $A_f(0,\Delta)$ and $A_f(\Delta,2\Delta)$ respectively. The orange coloured moon-edge represents the area $A_f(0,2\Delta)$. Due to the superaditivity property of a convex function, the dotted orange line always lies above the dotted blue lines making the blue filled area only a subset of the orange filled one.}
        \label{MECompare}
    \end{figure}
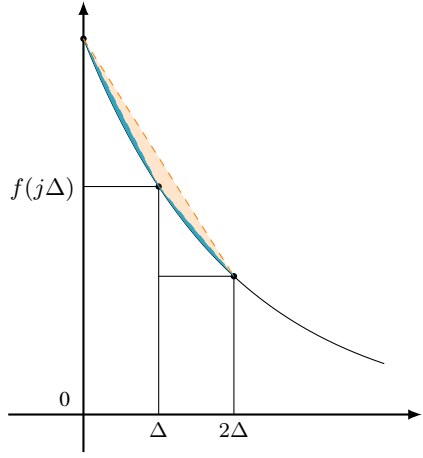

All that remains to do is to prove that $ M_f(\Delta) $ has a global minimum at the miminum possible value of $\Delta=\Delta_{0}(=2)$ for our theorem to hold. To this end, consider the simplest situation where the area of the moon-edges for $\Delta_0$ and $\Delta_2=2\Delta_0$ are compared. 
In accordance with Fig.~\ref{MECompare}, let $h_1(x)$ and $h_2(x)$ be defined on $[0,\Delta_0]$ and $[\Delta_0,2\Delta_0]$, respectively, both equal to $f(x)$ in the corresponding intervals.
Let $L_1(x)$ ($x\in[0,\Delta_0]$) and $L_2(x)$ ($x\in[\Delta_0,2\Delta_0]$) be the linear functions defined in terms of $h_1(x)$ and $h_2(x)$, respectively, in the way described in Lemma \ref{Lemma:convex_positive}.
Define $h(x)$ on $[0,2\Delta_0]$ as $h(x)=L_1(x)$ for $x\in[0,\Delta_0]$, and $h(x)=L_2(x)$ for $x\in[\Delta_0,2\Delta_0]$.
Let $L(x)$ ($x\in[0,2\Delta_0]$) be the linear function defined in terms of $h(x)$ in the way described in Lemma \ref{Lemma:convex_positive}.
Then
\begin{eqnarray}
\mathcal{A}_f(0,2\Delta_0)-[ \mathcal{A}_f(0,\Delta_0)+ \mathcal{A}_f(\Delta_0,2\Delta_0)]=\int_{0}^{2\Delta_0}dx[L(x)-h(x)].
\label{eq:Delta_2_minimum}
\end{eqnarray}
On the other hand, the second order derivative of $h(x)$ diverges to $+\infty$ at $x=\Delta_0$ and vanishes everywhere else.
Therefore $h(x)$ is a convex-down function since its second order derivative is always non-negative in the domain. 
Applying Lemma \ref{Lemma:convex_positive},  we see that Eq. (\ref{eq:Delta_2_minimum}) is always positive.
Apparently, the above discussion applies equally well when $0,\Delta_0,2\Delta_0$ are replaced by $i\Delta_0,(i+1)\Delta_0,(i+2)\Delta_0$, respectively, for any positive integer $i$.
Hence, we obtain
\begin{eqnarray}
M_f(2\Delta_0)&=&\sum_{i=0}^{\infty} \mathcal{A}_f(2i\Delta_0,2(i+1)\Delta_0)\nonumber\\
&>&\sum_{i=0}^{\infty}[\mathcal{A}_f(2i\Delta_0,(2i+1)\Delta_0)+\mathcal{A}_f((2i+1)\Delta_0,2i\Delta_0)]\nonumber\\
&=&M_f(\Delta_0).
\end{eqnarray}

In fact, using the same spirit of analysis, it follows that for any $ j \in \mathbb{N}_0, k \in \mathbb{N}$ $$ A_f(jk\Delta_0,(j+1)k\Delta_0) \geq \sum_{i=0}^{k-1} \mathcal{A}_f((jk+i)\Delta_0,(jk+i+1)\Delta_0) $$
Summing over $j$ in the above expression we have that for any $ k \in \mathbb{Z}_+: M_f(k\Delta_0)\geq M_f(\Delta_0)$,
i.e., $M_f(\Delta)\geq M_f(2)$ where $\Delta\in 2\mathbb{Z}_+$.
 This concludes the proof of Theorem 2. $\Box$

\section{:  MBQC on the surface code state}\label{SC-MBQC}

\begin{figure}
\begin{center}
\includegraphics[width=7cm]{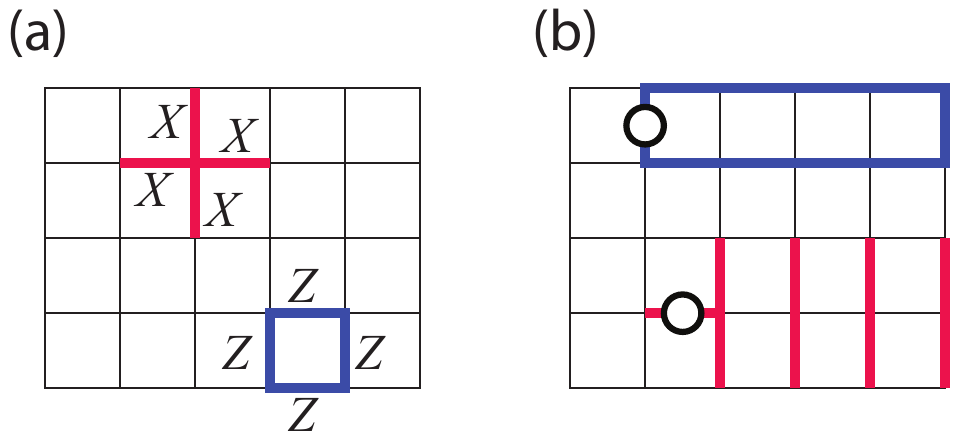}
\caption{\label{SC}Surface code state; qubits live on the edges. (a) The site and plaquette stabilizer generators. (b) The order parameter operators facilitating gate operation. They are in the stabilizer of the surface code state, and are the counterparts of the operators ${\cal{K}}(d)$ of the 1D cluster case. The qubits marked by circles represent the locations of non-Pauli measurements associated with the order parameters shown. }
\end{center}
\end{figure}

Fig.~\ref{SC} shows the surface code as computational resource for MBQC. It is invoked in the `Conclusion and Discussion' section in the main text, to illustrate Conjecture 2. The string order parameters $\langle {\cal{K}}_\geq\rangle$ are replaced by the stabilizer operators shown in Fig.~\ref{SC}b.

\end{appendices}

\end{document}